\documentclass[aps,pre,twocolumn,showpacs,superscriptaddress]{revtex4-1}
\usepackage{graphicx}
\usepackage{ulem}
\usepackage{color}
\usepackage{amsmath,amssymb}
\def\vector#1{\mbox{\boldmath $#1$}}

\begin{document}
\title{Phase diagram of the two-dimensional dipolar Heisenberg model with the Dzyaloshinskii-Moriya interaction and the Ising anisotropy}

\author{Hisato \surname{Komatsu}}
\email[Email address: ]{komatsu.hisato@nims.go.jp}

\affiliation{Research Center for Advanced Measurement and Characterization, National Institute for Materials Science, Tsukuba, Ibaraki 305-0047, Japan}

\author{Yoshihiko Nonomura}
\email[Email address: ]{nonomura.yoshihiko@nims.go.jp}

\affiliation{International Center for Materials Nanoarchitectonics,  National Institute for Materials Science,  Tsukuba, Ibaraki 305-0044, Japan}

\author{Masamichi Nishino}
\email[Email address: ]{nishino.masamichi@nims.go.jp} 

\affiliation{Research Center for Advanced Measurement and Characterization, National Institute for Materials Science, Tsukuba, Ibaraki 305-0047, Japan}

\affiliation{Elements Strategy Initiative Center for Magnetic Materials, National Institute for Materials Science, Tsukuba, Ibaraki 305-0047, Japan}

\begin{abstract}
We study phase transitions in the two-dimensional Heisenberg model with the Dzyaloshinskii-Moriya interaction, the Ising anisotropy ($\eta$), and the dipolar interaction under zero and finite magnetic fields ($H$). For three typical strengths (zero, weak, and strong) of the dipolar interaction, we present the $H$-$\eta$ phase diagrams by estimating order parameters for skyrmion-lattice and helical phases and in-plane magnetization by using a Monte Carlo method with an $O(N)$ algorithm. We find in the phase diagrams three types of skyrmion-lattice phases, i.e., two square lattices and a triangular lattice, helical phases with diagonal and vertical (or horizontal) stripes, canted ferromagnetic phase and polarized ferromagnetic phase. The effect of the dipolar interaction varies the types of the skyrmion and helical phases in a complex manner. The dipolar interaction also expands the regions of the ordered phases accompanying shifts of the phase boundaries to the positive $H$ and $\eta$ directions, and causes increase of the density of skyrmions and shortening of the pitch length (stripe width) of helical structures. We discuss the details of the features of the phase transitions.   
\end{abstract}
\maketitle

\section{Introduction }

Ultrathin magnetic films have attracted much attention for applications toward magnetic recording and their unique magnetic properties have drawn scientific interest~\cite{Heinrich,Bader}. 
The competition between magnetic anisotropies, short-range exchange and long-range dipolar interactions causes complex magnetic orderings such as spin-reorientation transitions between in-plane and out-of-plane magnetic phases and a variety of stripe patterns~\cite{Pappas,Allenspach,Ramchal,Won,Qiu}. 

The theoretical aspects of these phenomena have been often investigated by using the two-dimensional (2D) dipolar Ising~\cite{Booth,MacIsaac-Ising,Toloza,Rastelli06,Cannas,Pighin-Ising,Rastelli,Vindigni,Rizzi,Fonseca,Ruger,Horowitz,Bab,Leib,Komatsu1} or Heisenberg~\cite{Pescia,Moschel,Hucht,MacIsaac1,MacIsaac2,Bell,Santamaria,Rapini,Whitehead,Carubelli,Mol,Pighin2,Pighin,Mol2,Komatsu2} model with a magnetic anisotropy, and the phase diagrams with multiple stripe-ordered phases have been shown in several parameter regions. Reentrant transitions associated with planar ferro, stripe, and paramagnetic phases have also been presented~\cite{Komatsu1,Komatsu2}. 

The Dzyaloshinskii-Moriya (DM) interaction plays an important role in systems whose spatial reversal symmetry is broken, and it causes weak ferromagnetism or helimagnetism. Recently a topologically protected magnetic structure called (magnetic) skyrmion has been observed experimentally~\cite{Muhlbauer,Yu,Jonietz11,Yu11,Schulz12,Sampaio13,Jiang15,Boulle16}, and because of its potential applications to spintronics devices, physical properties of skyrmions have drawn much attention. For skyrmion-lattice phases, the competition between the DM and exchange interactions is important.

Skyrmion-lattice phases are more stabilized in 2D systems (thin films) than in 3D ones~\cite{Yu}. In 2D systems, the phase diagrams associated with skyrmion-lattice phases have been actively investigated by theoretical and computational methods with the use of the Heisenberg model with the DM interaction and with and without several anisotropies, and their phase diagrams in specific parameter regions have been shown~\cite{Yi,Kwon,Banerjee, Lin,Rowland,Nishikawa,Bernard}. 
However, studies on the effect of the dipolar interaction on the models~\cite{Kwon,Bernard} have hardly been performed and have not been well understood because of numerical difficulty for treating long-range interactions. 

In the present paper, we study the 2D classical Heisenberg model with the DM interaction, Ising anisotropy, and dipolar interaction under zero and finite fields. 
We show the field vs. Ising anisotropy ($H$-$\eta$) phase diagrams for typical three cases of the strength of the dipolar interaction, i.e., zero, weak, and strong interactions. 

We perform a direct simulation of order parameters in the model by using a Monte Carlo method. The most serious difficulty in numerical computation is $O(N^2)$ ($N$: the total number of spins) computational time originating from the fully-connected $O(N^2)$ long-range interactions, and we adopt the stochastic cut-off (SCO) $O(N)$ method~\cite{Sasaki} to reduce the computational cost.

The $H$-$\eta$ phase diagram in the ground state without the dipolar interaction was studied using variational approaches for equivalent continuum models. 
It was shown that in the case of small DM interaction, for small $|\eta|$, triangular skyrmion-lattice and helical phases exist at high and low (including zero) fields, respectively, which are located between the canted ferromagnetic (in-plane magnetic) phase at small $\eta$ and polarized ferromagnetic phase at large $\eta$~\cite{Kwon,Banerjee}. 

Afterward, Lin et al.\ presented that a square skyrmion-lattice phase can exist at finite fields and at a specific small region of $\eta$~\cite{Lin}. 
This phase is located at higher fields than the helical phase and between the canted ferromagnetic phase at smaller $\eta$ and the triangular skyrmion-lattice phase at larger $\eta$. 

In the present paper, we report the following new findings. 
Without the dipolar interaction, when the DM interaction is the same order of the exchange interaction, the square skyrmion-lattice phase exists in a wider region of the phase diagram including zero field. 
The dipolar interaction expands the regions of the ordered states whose boundaries shift to the positive $H$ and $\eta$ directions, increases the density of skyrmions, and shortens the pitch length (stripe width) of helical structures. 
The dipolar interaction induces another type of square skyrmion-lattice phase with a different alignment, and in the weak strength, a reentrance of a vertical (or horizontal) helical phase through a diagonal helical phase takes place with increasing $\eta$. 
These complex situations cause five kinds of triple points in the phase diagrams. 

The outline of the present paper is organized as follows. 
The model and method are explained in Sec.~\ref{model}. 
In Sec.~\ref{results}, the order parameters and magnetic structures are analyzed and the phase diagrams are presented. After the overview of this long section in Sec.~\ref{overview}, discussions about the characteristics of the model with no, weak, and strong dipolar interactions are given in Secs.~\ref{no-dipole}, \ref{weak-dipole}, and \ref{strong-dipole}, respectively.  A variational analysis is performed in Sec.~\ref{helical} for understanding the mechanism of the reentrant transition of the helical phase. 
The summary is given in Sec.~\ref{summary}.

\begin{figure}
\begin{center}
\includegraphics[width = 5cm]{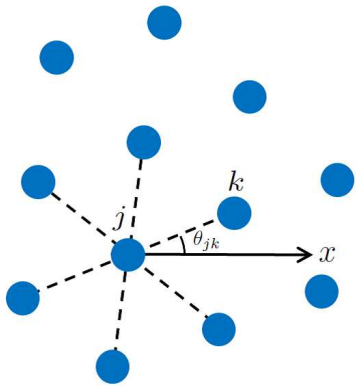} \\
\end{center}
\caption{Definition of $\theta_{jk}$ in Eq.~(\ref{Psi}). Blue circles denote the centers of skyrmions. }
\label{angle}
\end{figure}

\section{Model and method \label{model}}

In this study, we consider the system on a square lattice in the $xy$ plane composed of classical Heisenberg spins $\vector{S} _{i}$ with $|\vector{S} _{i}|=1$ at the position $ \vector{r}_i$ represented by the following Hamiltonian: 
\begin{eqnarray}
{\cal H} & = & - J \sum _{\left<i,j \right> } \vector{S} _{i} \cdot \vector{S} _{j} \\
&&- D \sum _{i} \left\{\vector{e}_x\cdot  (\vector{S}_{i} \times \vector{S}_{i +\hat{x}} ) + \vector{e}_y\cdot (\vector{S}_{i} \times \vector{S}_{{i} +\hat{y}} ) \right\} \nonumber \\
& & + D' \sum _{i < j  } \left\{ \frac{\vector{S} _{i} \cdot \vector{S} _{j} }{r_{ij}^3} - \frac{ 3(\vector{S} _{i} \cdot \vector{r}_{ij} ) (\vector{S} _{j} \cdot  \vector{r}_{ij} ) } {r_{ij} ^5} \right\}  \nonumber \\
&&- \eta \sum_{i} \left( S _{i}^z \right) ^2 - H \sum _{i} S _{i}^z ,   \nonumber
\label{Hamiltonian}
\end{eqnarray}
with the coupling constants $J$ for the nearest-neighbor ferromagnetic interaction, $D$ for the DM one, $D'$ for the dipolar one, $\eta$ for the Ising anisotropy, and $H$ for the magnetic field along the $z$ axis. 
Vector $\vector{r}_{ij}$ is defined as $\vector{r}_{ij} \equiv \vector{r}_j- \vector{r}_i$, $\vector{e}_x$ and $\vector{e}_y$ are unit vectors in the $x$ and $y$ directions, respectively, and site $i+\hat{x}$ ($i+\hat{y}$) is the nearest neighbor one of site $i$ in the positive $x$ ($y$) direction. We consider both positive and negative ($XY$ anisotropy) values for $\eta$. 
Here we fix the coupling constants of short-range interactions as $J=D=1$. 

A previous study of the model for $J=D=1$ and $\eta=0$ without the dipolar interaction by Nishikawa et al.~\cite{Nishikawa} showed that a triangular skyrmion-lattice phase exists and the skyrmion lattice does not have the long-range positional order but has the long-range orientational order. 
In the present study, to identify the triangular and square skyrmion-lattice phases, we introduce the orientational order parameter which detects the $n$th rotational symmetry ($C_n$) as follows. 
First we detect a domain in which $S_{i} ^z < S_{\mathrm{th}} \le 0$ for $H \ge 0$ are fulfilled, and regard this domain as one skyrmion. 
Then, we define its position as the mean value of the positions of the spins inside the domain:
\begin{equation}
\vector{x} _k = \frac{ \sum _{ \vector{r}_i \in {\cal S}_k } \vector{r}_i }{ \sum _{ \vector{r}_i \in {\cal S}_k } 1 }  ,
 \label{xs}
\end{equation}
where ${\cal S}_k$ is the $k$th domain. 
For weak fields, down-spin regions are accidentally generated by thermal fluctuations between skyrmions. These down spins connect the skyrmions momentarily (Fig. \ref{snap2}(b)). To detect skyrmions precisely, we tune the cutoff value $S_{\mathrm{th}}$ depending on $H$. We take $S_{\mathrm{th}}$ smaller for weak fields as $S_{\mathrm{th}}=-0.5$ for $H \le 0.2$, while $S_{\mathrm{th}}=0$ for $H > 0.2$. 

The orientational order parameter is given as  
\begin{equation}
\Psi _n \equiv \frac{1}{N_s} \sum _j \frac{\sum _{k \in \partial j} e^{ i n\theta _{jk} } }{ \sum _{k \in \partial j} 1 }, 
\label{Psi}
\end{equation}
where $N_s$ denotes the number of skyrmions, $\theta _{jk}$ is the angle between 
vector $\vector{x} _k -\vector{x} _j$ and the $x$ axis, and $\partial j$ is the set of skyrmions adjacent to the $j$th one (Fig.~\ref{angle}). 
For the triangular and square skyrmion lattices, $n=6$ and $8$ are taken, respectively.  We give below the reason why we take $n=8$ instead of $n=4$ to detect the square lattice. 

We judge the adjacent skyrmions by the Delaunay triangulation~\cite{Shewchuk}. 
When the lattice is slightly distorted from the perfect square lattice, 
about half of the pairs of the next-nearest-neighbor skyrmions are regarded as adjacent ones, and this contribution to $\Psi _4$ partially cancels that from the pairs of nearest-neighbor skyrmions. On the other hand, when $\Psi_8$ is considered, the above two contributions have the same sign and such cancellation does not take place. We confirmed that $\Psi _8$ works as the order parameter of the square skyrmion-lattice phase. 
For skyrmion-lattice phases, we also apply the local chirality~\cite{Yi} commonly used for the detection of them: 
\begin{equation}
\chi = \frac{1}{8 \pi} \sum _{i} \vector{S} _{i} \cdot \left( \vector{S} _{i+\hat{x}} \times \vector{S} _{i + \hat{y} } + \vector{S}_{i - \hat{x} } \times \vector{S} _{i-\hat{y}} \right) .
 \label{chi}
\end{equation}
In the model (1), $\chi$ takes a negative value for skyrmion-lattice phases. 
It should be noted that $\chi=0$ by definition for the configuration with skyrmion and anti-skyrmion pairs at $H=0$ (Fig. \ref{snap2}(b)). Furthermore, $\chi$ detects skyrmions without orientational order, i.e., a liquid state. Therefore, we use $\chi$ to estimate the upper limit of skyrmion-lattice phases at high fields. 

In the helical phase, a stripe pattern of up and down spins is formed. 
To detect the helical phase, we measure another type of orientational order parameters~\cite{Whitehead} defined as
\begin{equation}
O_{\rm hv} = \left| \frac{n_{\rm h} - n_{\rm v}}{n_{\rm h} + n_{\rm v}} \right|  \;\; {\rm and}  \;\; \ O_{\rm d} = \left| \frac{n_{{\rm d}1} - n_{{\rm d}2}}{n_{{\rm d}1} + n_{{\rm d}2}} \right| ,
\label{Ohv}
\end{equation}
with 
\begin{eqnarray}
n_{\rm h} & = & \sum _{i} \left( 1 -  \vector{S}_i \cdot \vector{S}_{i+\hat{x}} \right),  \label{nh} \\
n_{\rm v} & = &\sum _{i} \left( 1 -  \vector{S}_i \cdot \vector{S}_{i+\hat{y}} \right), \label{nv} \\
n_{{\rm d}1} & = & \sum _{i} \left( 1 -  \vector{S}_i \cdot \vector{S}_{i+\hat{x}+\hat{y}} \right),  \label{ns1} \\
{\rm and} \;\;\;    n_{{\rm d}2} & = &\sum _{i} \left( 1 -  \vector{S}_i \cdot \vector{S}_{i+\hat{x}-\hat{y}} \right).  \label{ns2} 
\end{eqnarray}
Both of these order parameters correspond to the breaking of the symmetry under the $\frac{\pi}{2}$-rotation. The value of $O_{\rm hv}$ is the maximum when [1,0] or [0,1] helical structure is formed, whereas that of $O_{\rm d}$ is the maximum when [1,1] or [1,$-$1] helical structure is formed.

We also measure the uniform in-plane magnetization 
\begin{equation}
M_{xy} = \frac{1}{N} \sqrt{ \left( \sum _i S_i ^x  \right) ^2 + \left( \sum _i S_i ^y  \right) ^2 }
\label{Mxy}
\end{equation}
with the total number of spins $N$, and it has a nonzero value in the low-temperature phase when $\eta$ is sufficiently small. 

\begin{figure}
\begin{center}
\includegraphics[width = 7.6cm]{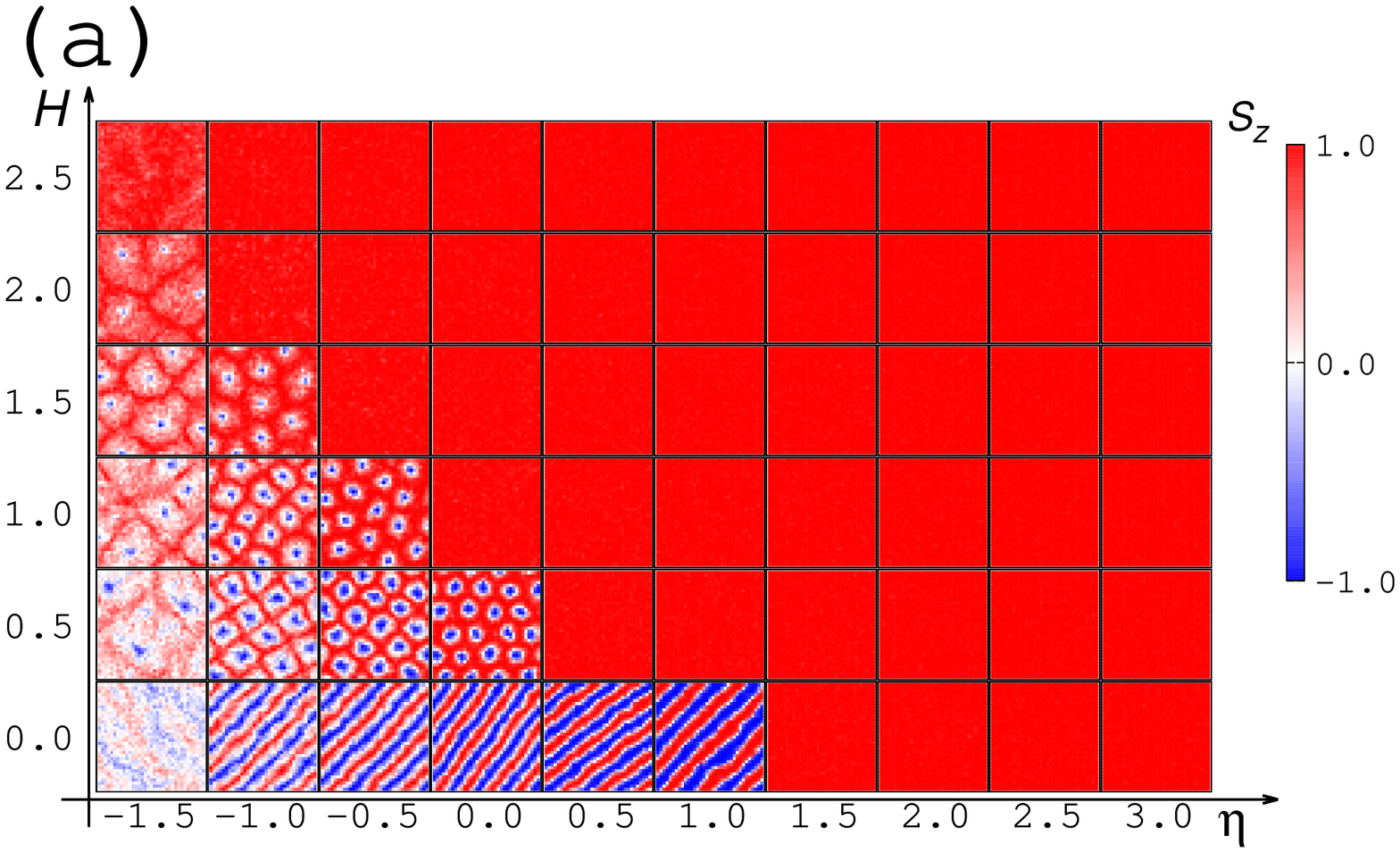} \\
\includegraphics[width = 7.6cm]{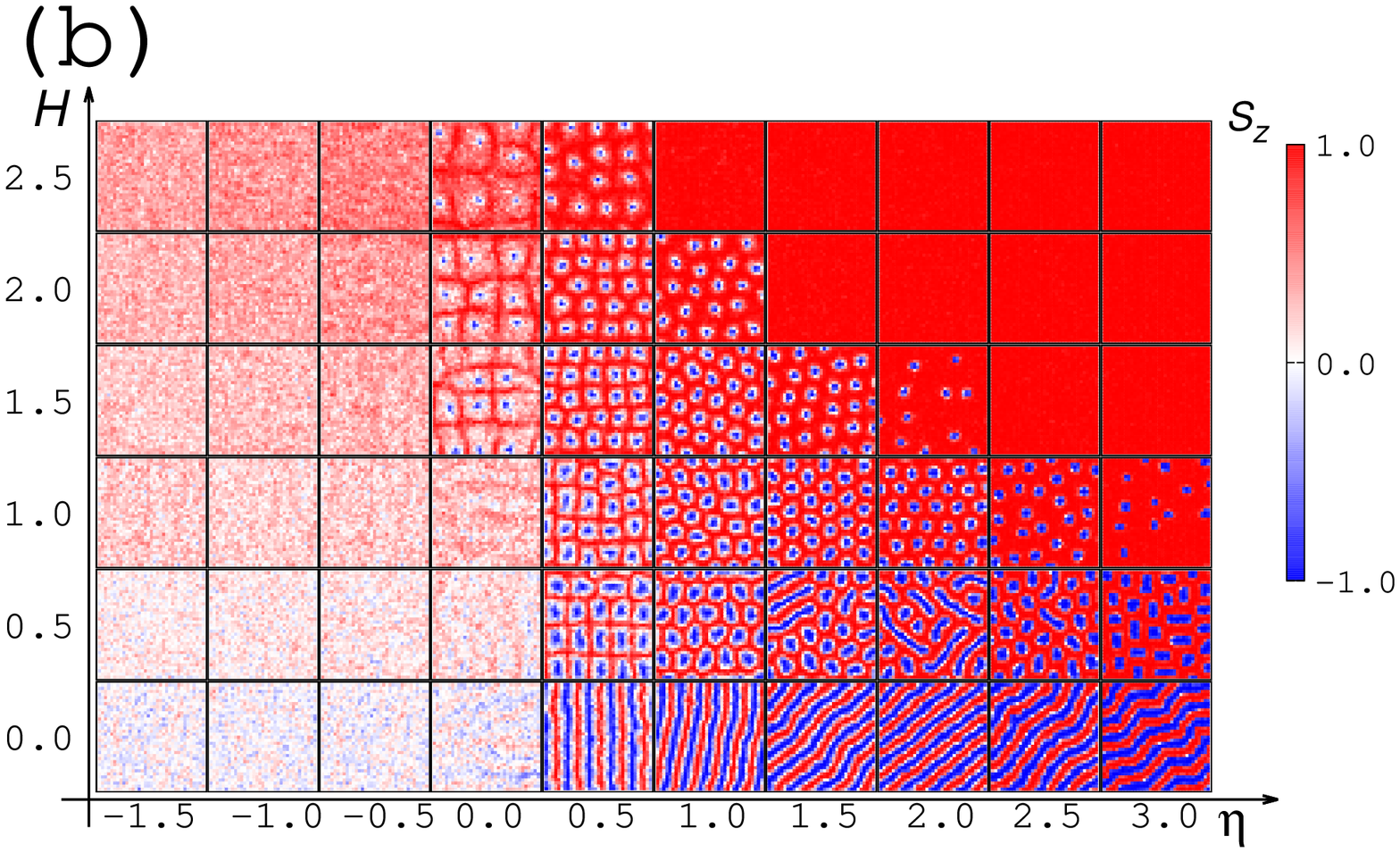} \\
\includegraphics[width = 7.6cm]{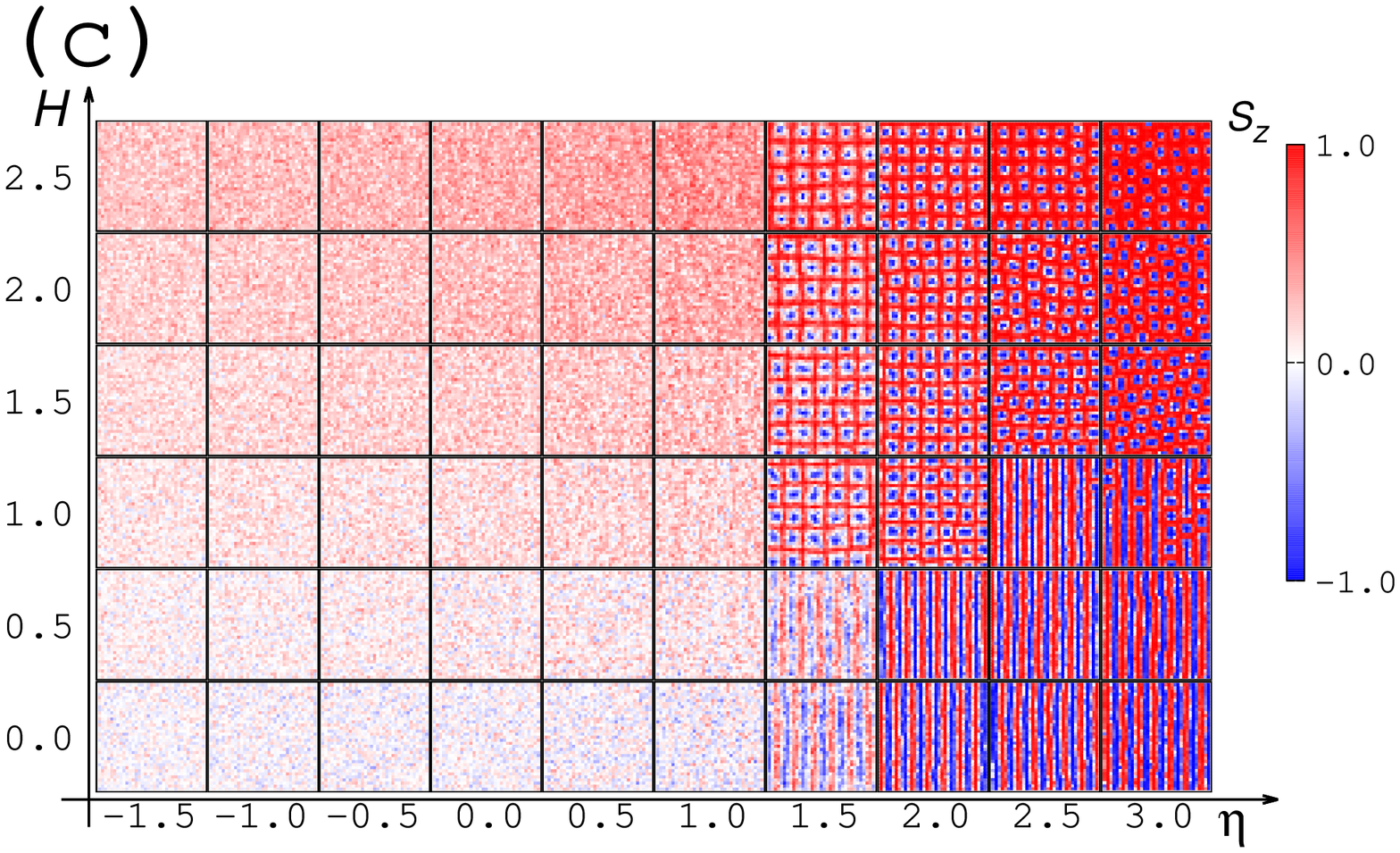}  \\
\end{center}
\caption{Snapshots of spin structures (red-and-blue contour plot of the $S^{z}$ element) for representative sets of $(H,\eta)$ for $L=36$ at $T=0.1$ for (a) $D'=0$, (b) $D'=0.3$, and (c) $D'=0.6$. }
\label{snap1}
\end{figure}

In Monte Carlo (MC) simulations of the present study, we use the stochastic cut-off (SCO) $O(N)$ method~\cite{Sasaki}, which reduces the computational cost for dipolar systems. Each Monte Carlo run for a set of $H$ and $\eta$ starts from a uniformly random state at a high temperature. 
For $\eta \ge 3.0$, the system is cooled from $T=0.34$ to $T=0.16$ at intervals of $T=0.02$ and further cooled to $T=0.1$, and for $\eta < 3.0$, it is cooled from $T=0.6$ to $T=0.1$ at intervals of $T=0.05$. At each temperature, 400,000 MC steps are used for the measurement after 100,000 MC steps for the equilibration. 
The order parameters are estimated by averaging over four independent simulations for the system size $L=84$ ($N=L^2$), and the error bar is estimated by $\pm \sigma /\sqrt{4}$, where $\sigma$ is the sample standard deviation. In Figs.~\ref{snap1} and \ref{snap2}, snapshots of spin structures for $L=36$ are shown, which are qualitatively the same as those for $L=84$. 

\begin{figure}
\begin{center}
\includegraphics[width = 7cm]{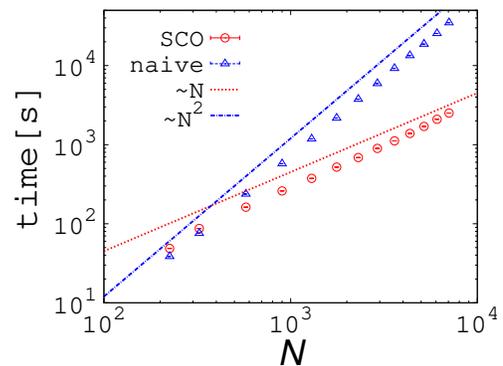} \\
\end{center}
\caption{Computational time [s] as a function of $N$ by the SCO and naive MC methods. Red dotted and blue dash-dotted lines are guide for eyes for $O(N)$ and 
$O(N^2)$ dependences, respectively.  }
\label{benchmark}
\end{figure}

We show a benchmark to compare the SCO method and a naive MC method in Fig.~\ref{benchmark}. The computational time [s] for 100,000 MCS is plotted as a function of $N$ for the parameters $D'=0.6$, $H=1.5$, and $\eta=2.0$.  
We confirm that the SCO method costs $O(N)$ computational time and is more efficient than the naive MC method which costs $O(N^2)$ computational time.

\begin{figure}
\begin{center}
\includegraphics[width = 7.5cm]{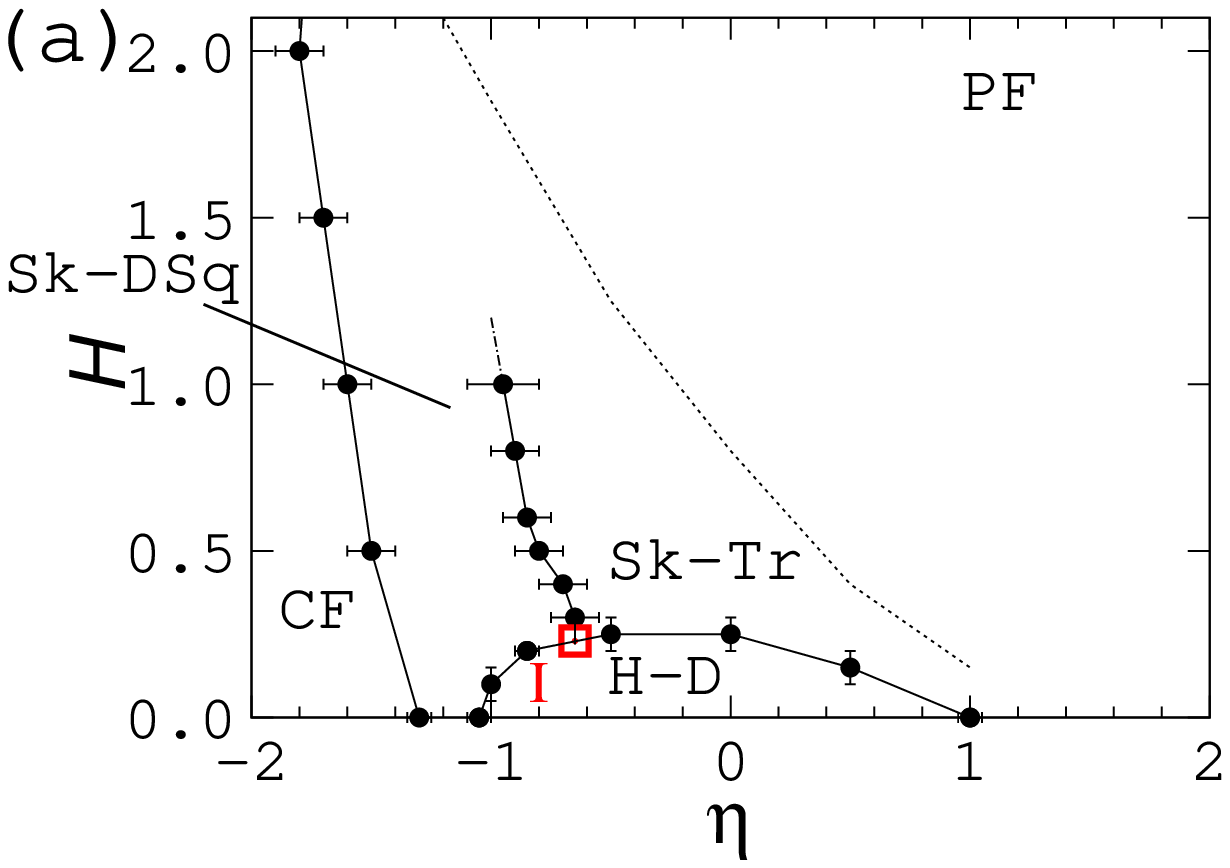} \\
\includegraphics[width = 7.5cm]{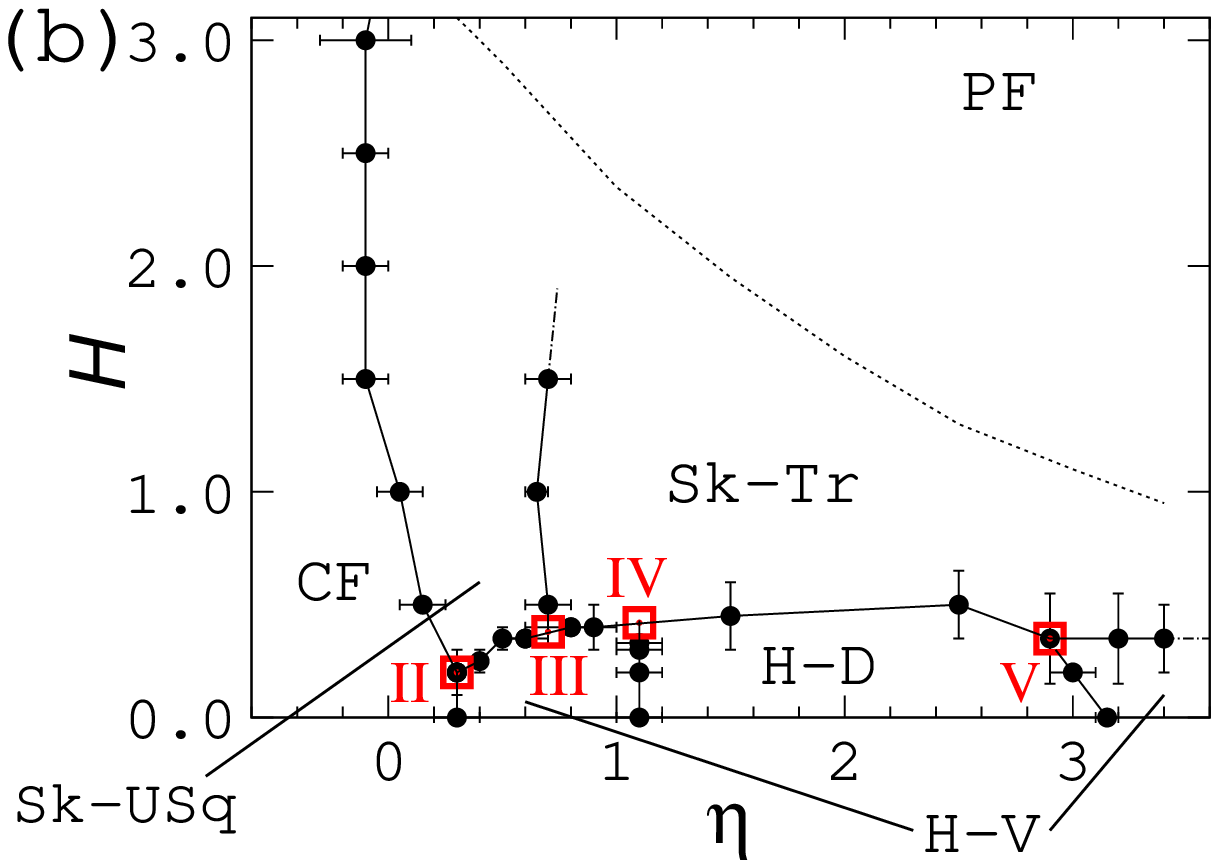} \\
\includegraphics[width = 7.5cm]{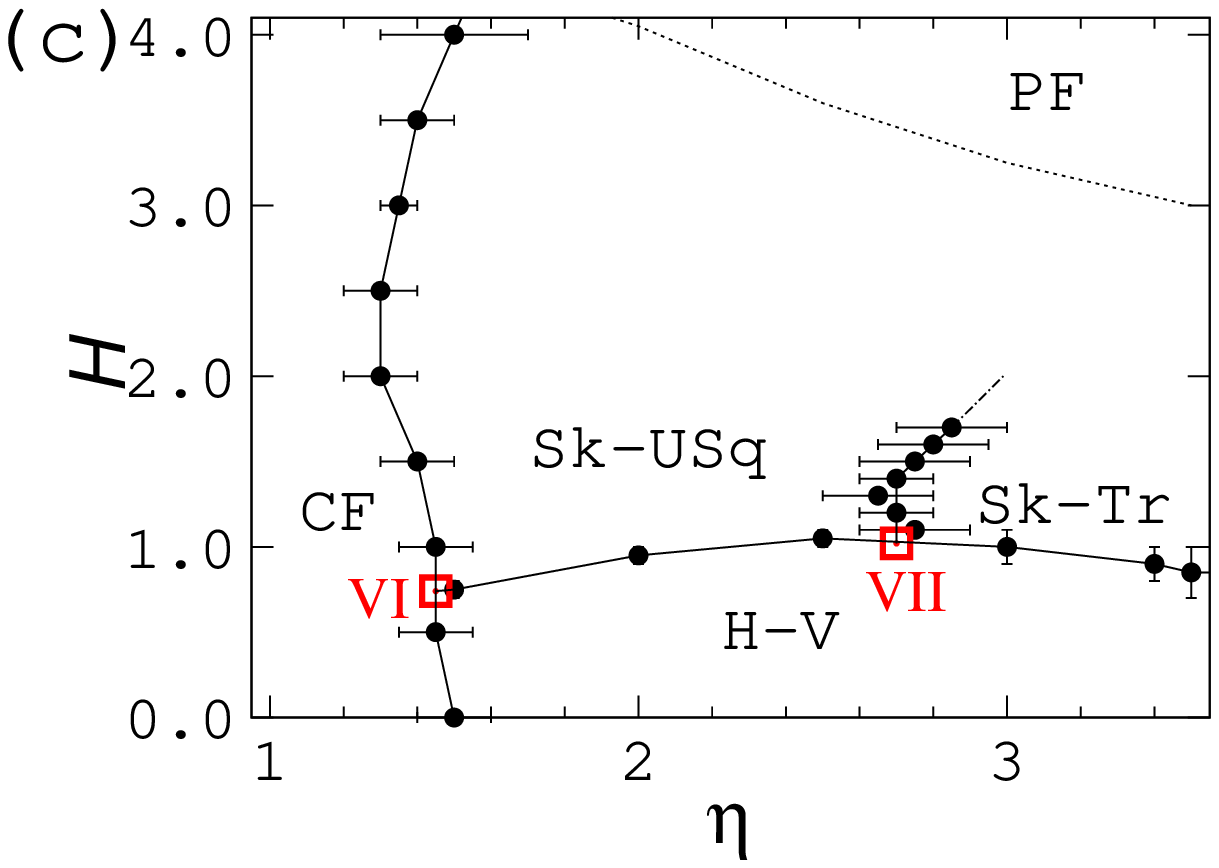} \\
\end{center}
\caption{Phase diagrams for (a) $D'=0$, (b) $D'=0.3$, and (c) $D'=0.6$ at $T=0.1$. 
The phase boundaries were estimated by the data for $L=84$. 
There exist three kinds of skyrmion phases, i.e., triangular (Sk-Tr), diagonal square (Sk-DSq), and upright square (Sk-USq) phases, and two kinds of helical phases, i.e., diagonal (H-D) and vertical (H-V) helical phases. Canted ferromagnetic (CF) phase exists at smaller $\eta$. Polarized ferromagnetic (PF) phase exits at larger $\eta$. Triple points I-VII are drawn by open squares. Dash-dotted lines stand for tentative boundaries between the square and triangular skyrmion-lattice phases at high fields, at which the boundaries are difficult to determine precisely because of small values of the order parameters $\Psi_8$ and $\Psi_6$.}
\label{PD}
\end{figure}

\section{Results \label{results} }

\subsection{Overview on magnetic structures}
\label{overview}
\begin{figure}
\begin{center}
\includegraphics[width = 9.0cm]{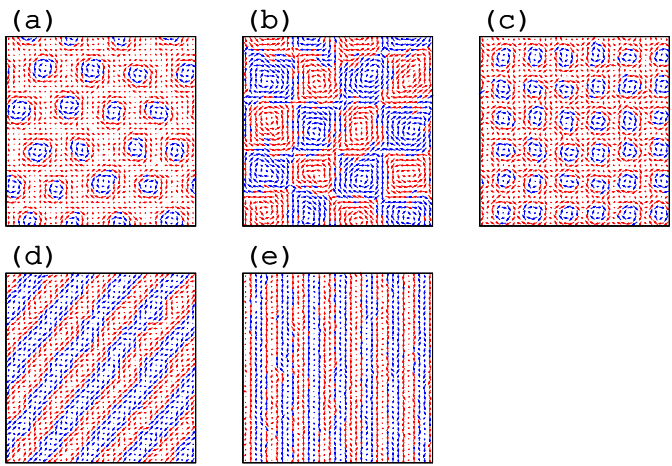} 
\end{center}
\caption{Typical snapshots of the ordered phases for $L=36$ at $T=0.1$.  
(a) The triangular skyrmion-lattice phase at $D'=0$, $H=0.5$ and $\eta = 0$. (b) The diagonal square skyrmion-lattice phase at $D'=0$, $H=0$ and $\eta = -1.2$. 
(c) The upright square skyrmion-lattice phase at $D'=0.6$, $H=1.5$ and $\eta = 2$. 
(d) The diagonal helical phase at $D'=0$, $H=0$ and $\eta =0$. 
(e) The vertical helical phase at $D'=0.6$, $H=0$ and $\eta = 2$. 
In each snapshot, the directions of the vectors indicate $(S^x , S^y)$ of each spin, and the colors represent the signs of $S^z$. Namely, red and blue vectors denote $S^z > 0$ and $S^z < 0$, respectively.}
\label{snap2}
\end{figure}

Main results in the present article are summarized in the $H$-$\eta$ phase diagrams at $T=0.1$ for no ($D'=0.0$), weak ($D'=0.3$), and strong ($D'=0.6$) dipolar interactions in Figs.~\ref{PD}  (a), \ref{PD} (b), and \ref{PD} (c), respectively. 
Snapshots of representative patterns of spin configurations in the $H$-$\eta$ space are displayed in Fig.~\ref{snap1} for (a) $D'=0.0$, (b) $D'=0.3$, and (c) $D'=0.6$. 
Some enlarged snapshots for typical magnetic orderings are also shown in Fig.~\ref{snap2}. 
Around the boundary of two phases, there is a region in which the two order parameters are of the same order. The error bar of the phase boundary is defined so as to include this region. The phase boundary is defined as the middle point of the error bars.

At zero and low $H$ for $D'=0$, the helical phase exists at small $|\eta|$. 
The strength of $D'$ changes the direction of the stripe structure of the helical phase. For zero and small $D'$, [1,1] or [1,$-$1] helical structure is formed, while for large $D'$, [1,0] or [0,1] helical one is formed. 
We call the former and latter phases diagonal helical phase and vertical helical one (vertical and horizontal ones are equivalent), respectively. 
For intermediate $D'$, a complex situation takes place, i.e., the type of the helical structure depends on $\eta$. 

Skyrmion-lattice phases exist regardless of the value of $D'$. 
However, the types of the lattice vary depending on $D'$. 
We find a triangular skyrmion-lattice phase (Fig.~\ref{snap2} (a)), and two kinds of square skyrmion-lattice phases, in which the nearest-neighbor skyrmion aligns in the [1,1] or [1,$-$1] direction (call diagonal square skyrmion-lattice phase) as shown in Fig.~\ref{snap2} (b) or  [1,0] or [0,1] direction (call upright square skyrmion-lattice phase) as shown in Fig.~\ref{snap2} (c). 

The diagonal and upright square lattices are not discerned by the orientational order parameter $\Psi_8$ and are distinguished by the snapshots. 
The dotted lines stand for the vanishing lines of the local chirality, 
namely the vanishing lines of skyrmion structures~\cite{Lenov}. As explained above, there may exist the skyrmion liquid phase where skyrmions are stable 
but do not form lattices, and these lines can be regarded as the upper 
limit of the skyrmion-lattice phases. 

The short-range part of the dipolar interaction has been studied as a local anisotropy term~\cite{Kashuba}, which acts as an easy-plane term ($\eta<0$). The shift of the canted ferromagnetic phase to the positive $\eta$ direction by the dipolar interaction is qualitatively explained by this contribution, but the dipolar interaction qualitatively changes skyrmion-lattice and helical phases, which is owing to the non-local effect.

%

\begin{figure}
\begin{center}
\includegraphics[width = 6.5cm]{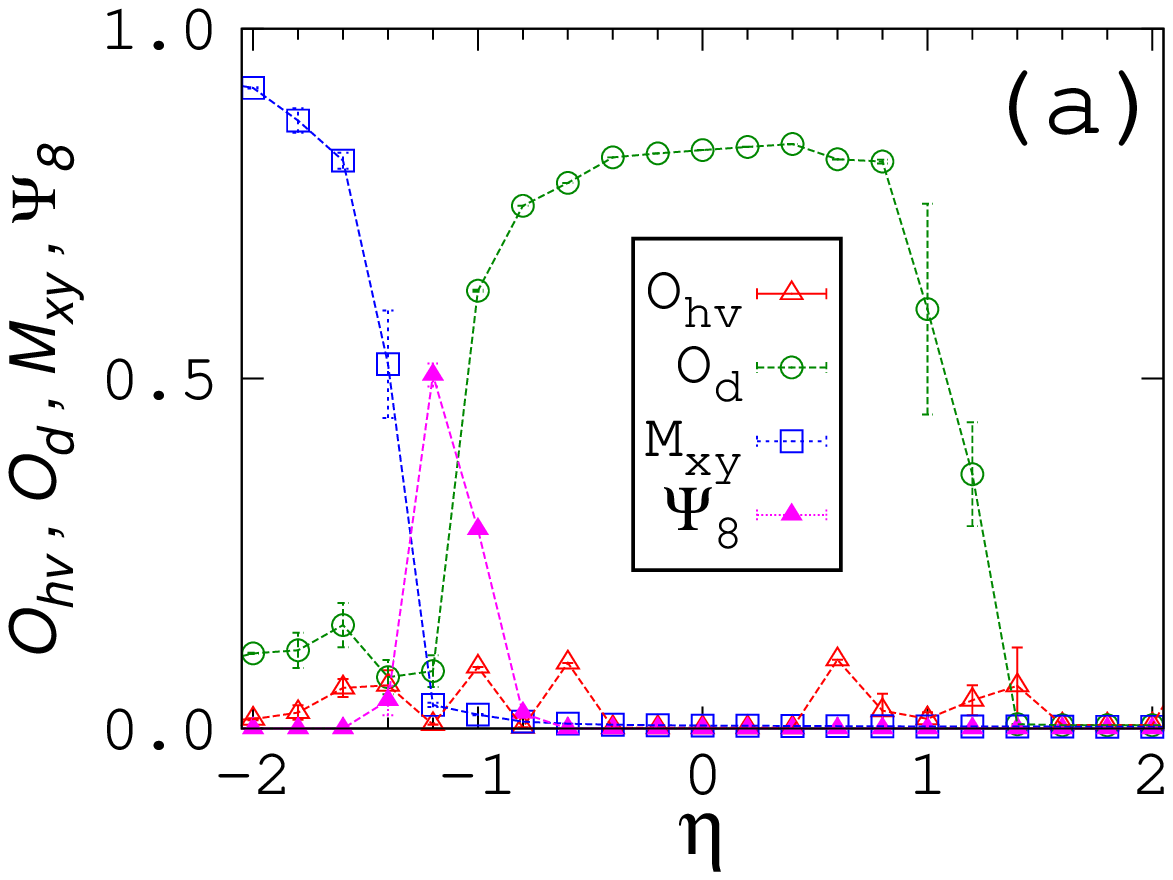} 
\hspace*{0.6cm}
\includegraphics[width = 7.25cm]{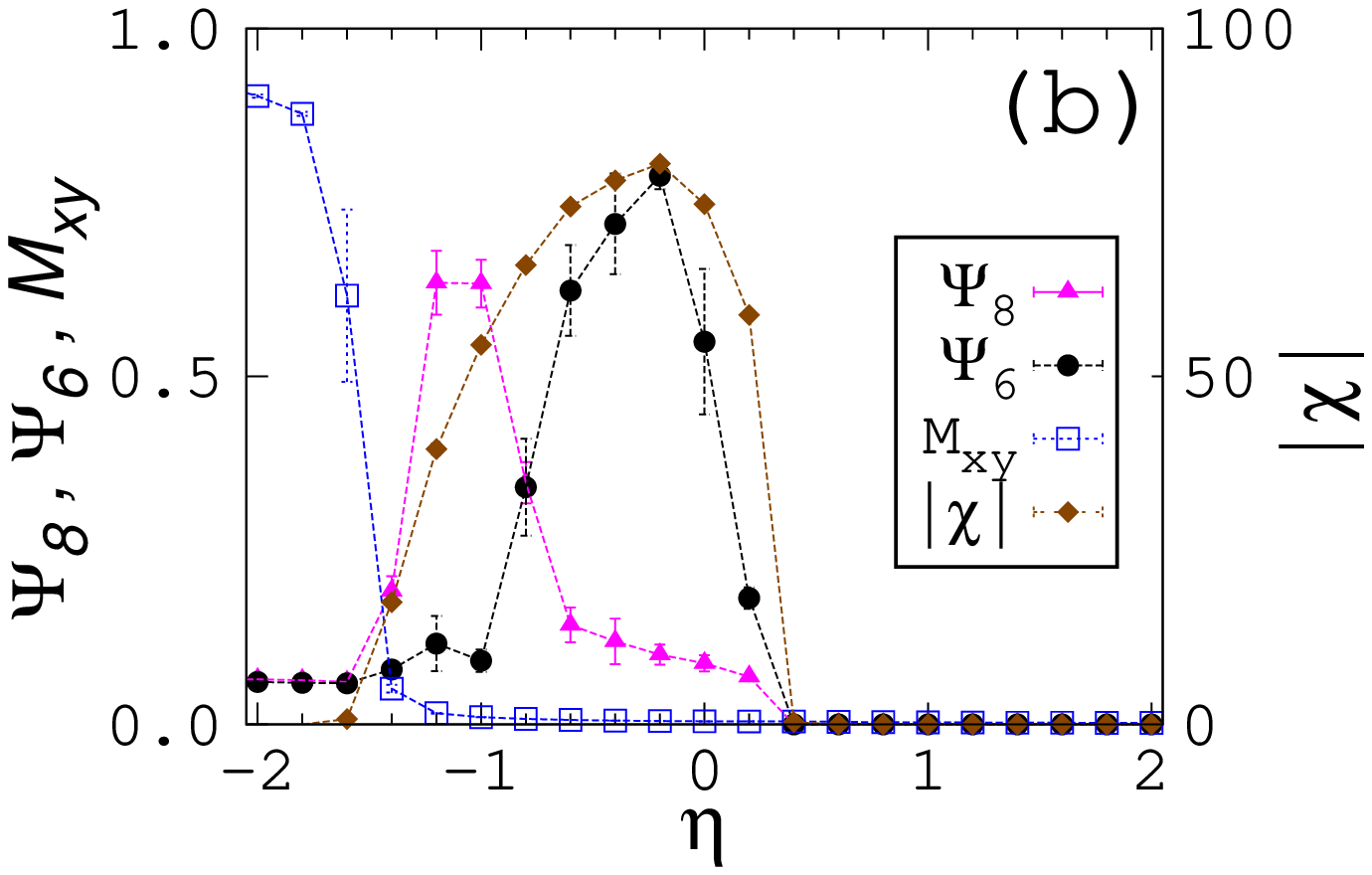} 
\end{center}
\caption{$\eta$ dependences of the order parameters at (a) $H=0$ and (b)  $H=0.5$ without the dipolar interaction $D'=0$ at $T=0.1$ for $L=84$. 
$O_{\rm hv}$, $O_{\rm d}$, $M_{xy}$, $\Psi  _8$ and $\Psi _6$ are order parameters for the vertical helical phase, the diagonal helical phase, the in-plane magnetization, the square skyrmion-lattice phase, and the triangular skyrmion-lattice phase, respectively. The local chirality $|\chi|$ signals existence of skyrmions, and can be regarded as a kind of order parameter.}
\label{O_dp00}
\end{figure}

\begin{figure}
\begin{center}
\includegraphics[width = 7.25cm]{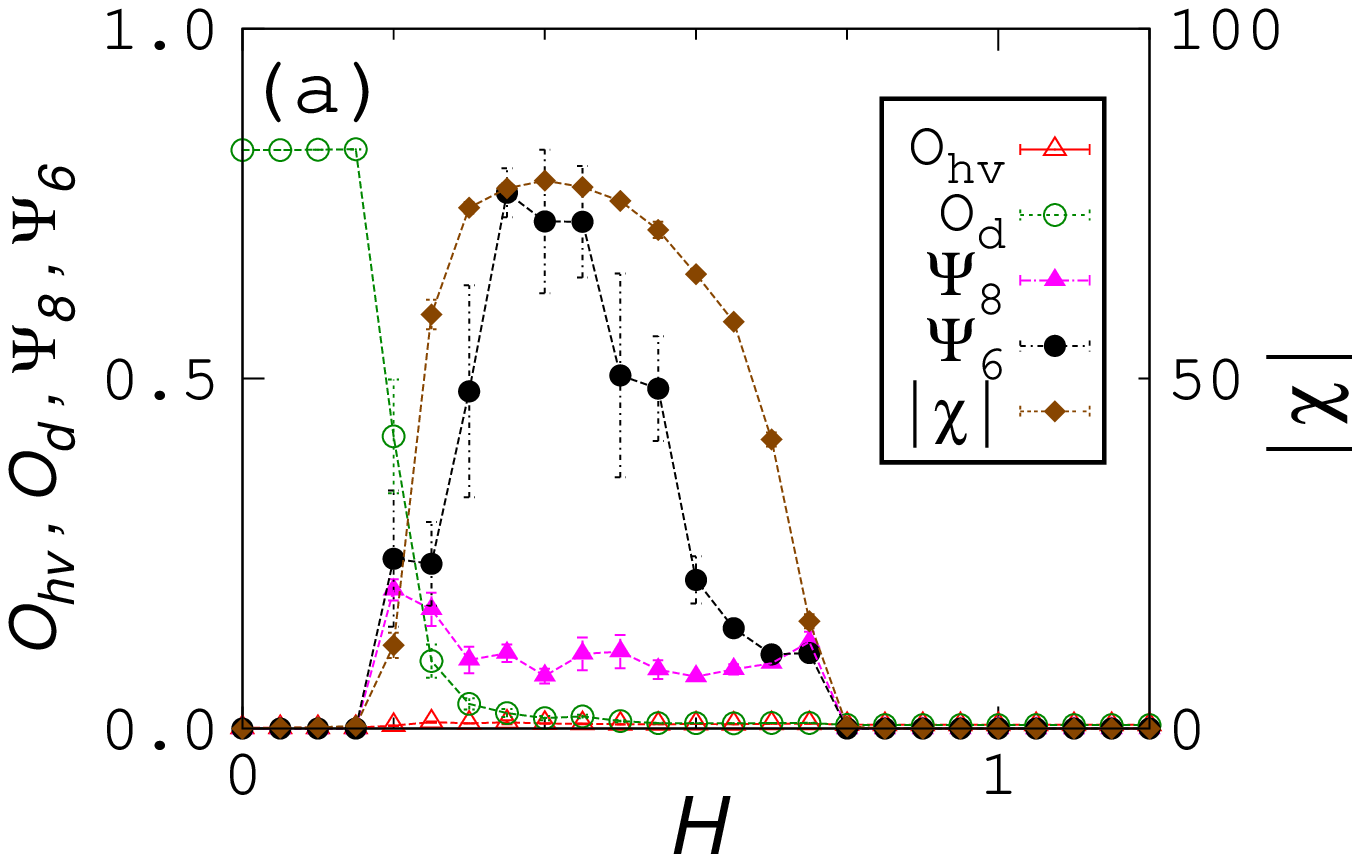} 
\includegraphics[width = 7.25cm]{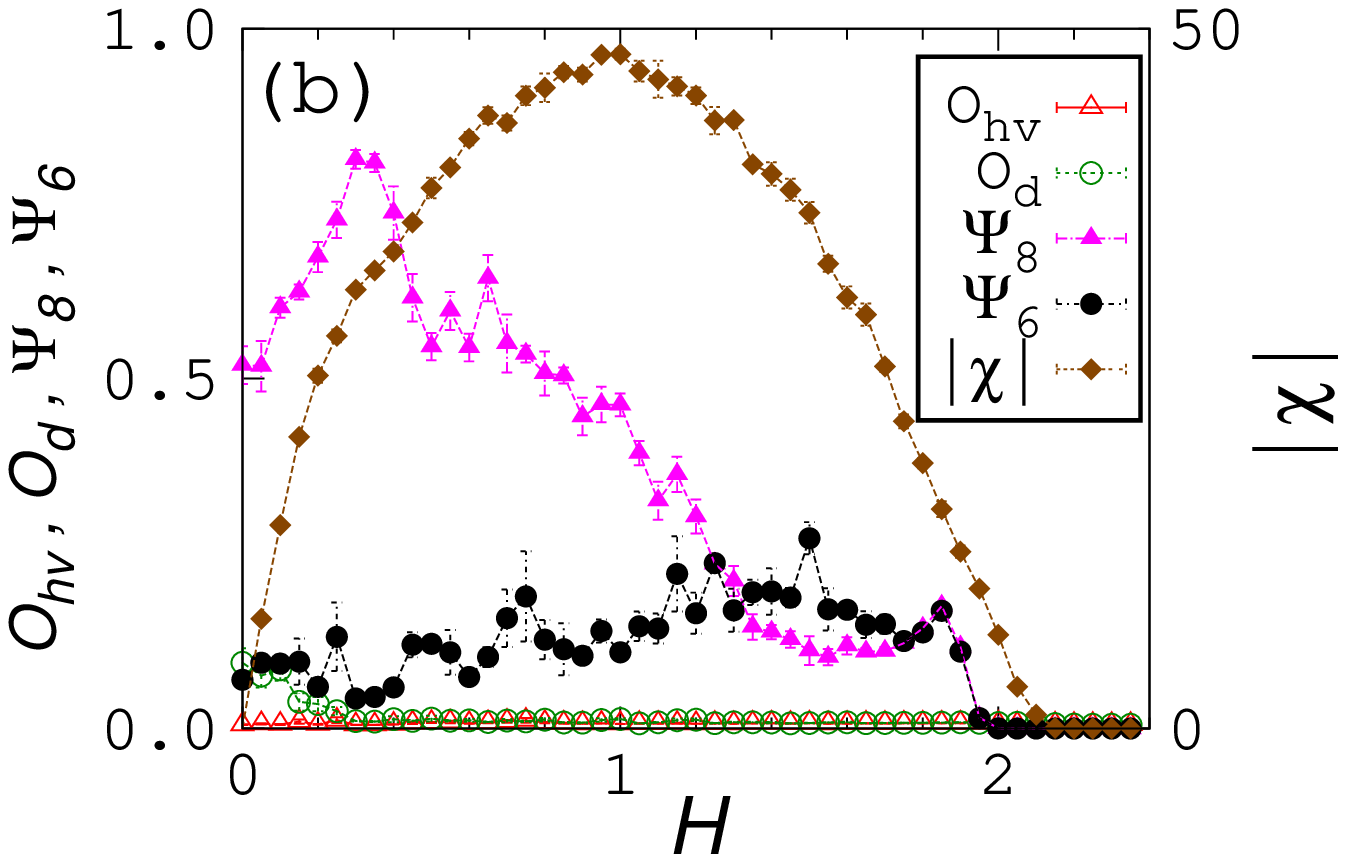} 
\end{center}
\caption{$H$ dependences of the order parameters at (a) $\eta=0$ and (b) $\eta=-1.2$ without the dipolar interaction $D'=0$ at $T=0.1$ for $L=84$. }
\label{O_dp00_Hdep}
\end{figure}

\subsection{The system without the dipolar interaction \label{sec_dp00} }
\label{no-dipole}

We investigate the model without the dipolar interaction $D' = 0$. 
The $\eta$-dependences of the order parameters at $H=0$ and $H=0.5$ are displayed in Figs.~\ref{O_dp00} (a) and \ref{O_dp00} (b), respectively (see also Fig.~\ref{PD} (a)). 
The $H$ dependences of the order parameters at $\eta=0$ and $\eta=-1.2$ are 
presented in Figs.~\ref{O_dp00_Hdep} (a) and \ref{O_dp00_Hdep} (b), respectively. 
As pointed out in the introduction, we find that $\chi=0$ at $H=0$, while $\Psi_8>0$ indicates the diagonal square skyrmion-lattice phase. 

At $\eta=0$, with increasing $H$ from zero, the diagonal helical phase changes to the triangular skyrmion-lattice one at $H=0.25\pm0.05$ as shown in Fig.~\ref{PD} (a), which is close to the transition point $H\sim0.28$ at $T=0.1$ estimated by eye in the $H-T$ phase diagram in Ref.~\cite{Nishikawa} for the model with $D'=0$ and $\eta=0$ in a 2D triangular lattice system. 

We find a characteristic structure in the phase diagram. 
At low fields ($0 \leq H \lesssim 0.25$), the diagonal square skyrmion-lattice phase appears between the canted ferromagnetic phase at smaller $\eta$ and the diagonal helical phase at larger $\eta$. 
At high fields, the diagonal helical phase disappears. Instead, the diagonal square skyrmion-lattice phase is expanded and the triangular skyrmion-lattice phase appears at larger $\eta$ than the diagonal square skyrmion-lattice phase. This suggests the existence of a triple point (call point I) at $(\eta,H) \simeq (-0.65,0.23)$ marked with an open square in Fig.~\ref{PD} (a), at which the diagonal square skyrmion-lattice phase, the triangular skyrmion-lattice phase, and the diagonal helical phase coexist. 

Kwon et al.\ studied the phase diagram for $J/D=3.3$ at $T=0$ using a variational approximation method, and showed that at zero and low fields, with increasing $\eta$, the canted ferromagnetic phase directly changes to the triangular skyrmion-lattice phase (no square skyrmion-lattice phase)~\cite{Kwon}. Banerjee et al.\ also showed a similar phase diagram to their phase diagram~\cite{Banerjee} for a small $D$.  
We calculated the order parameters and spin configurations for $J/D=3.3$ at $T=0.1$ and found that $\Psi_8$ and $\Psi_6$ are very small at $H=0$, and at high fields $\Psi_6$ appears, which is consistent with their results. 
We also notice that the shape of the region of the diagonal helical phase in the $H-\eta$ phase diagram is much more symmetric than the shape in the phase diagram by Kwon et al.\ or by Banerjee et al., in which the region is wider at smaller $\eta$. The cause of this difference is not clear, but it may be attributed to the difference in computational methods. Namely, Fig.~\ref{PD} is a result of the MC method, while their results are based on variational methods. 
Lin et al.\cite{Lin} presented a square skyrmion-lattice phase  (we call diagonal square skyrmion-lattice phase) in a narrow region between the canted ferromagnetic phase and the triangular skyrmion-lattice phase in a semiquantitative phase diagram for the equivalent continuum spin model. However, in their phase diagram any skyrmion-lattice phases do not appear at zero and low fields.  

We point out that when the value of $D$ is close to that of $J$, the diagonal square skyrmion-lattice phase is more stabilized and can exist at zero and low fields, and its region is not small especially at large fields. 

\begin{figure}
\begin{center}
\includegraphics[width = 6.5cm]{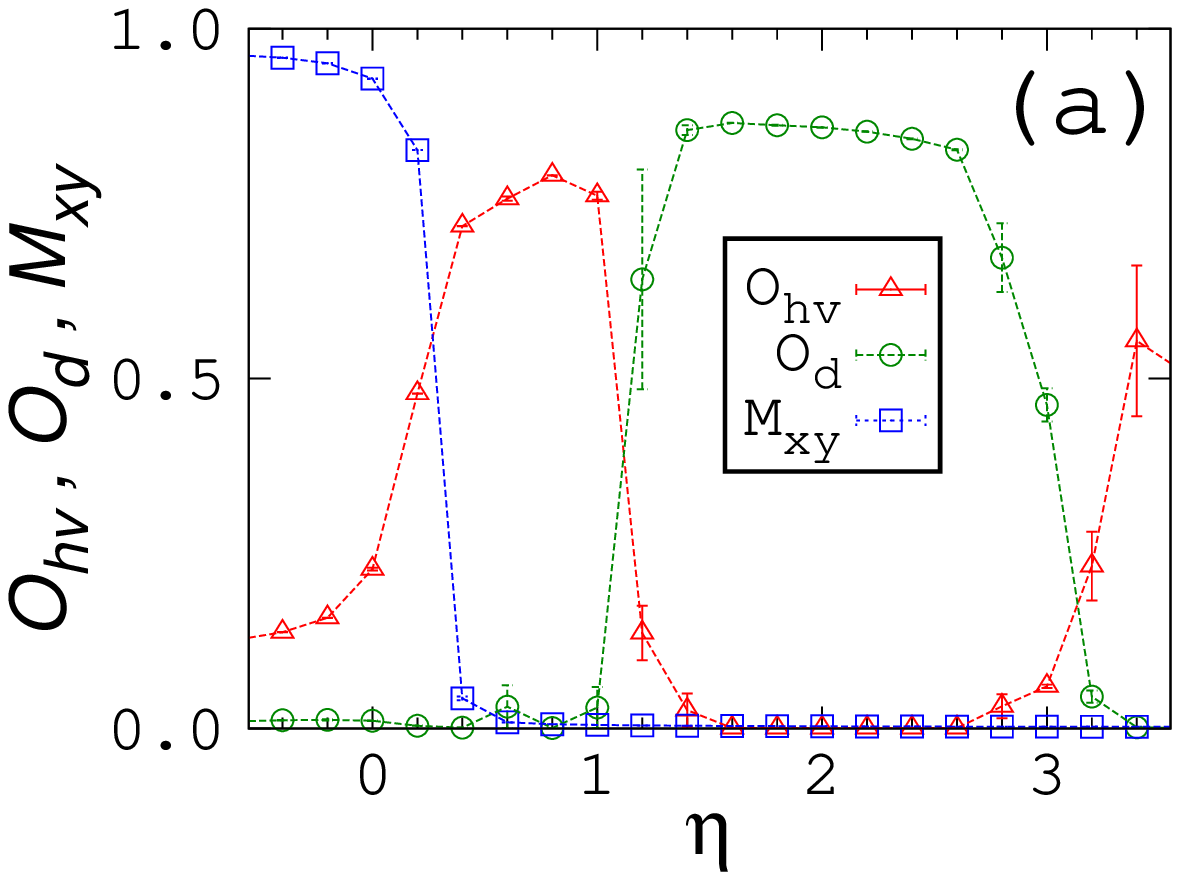} 
\hspace*{0.6cm}
\includegraphics[width = 7.25cm]{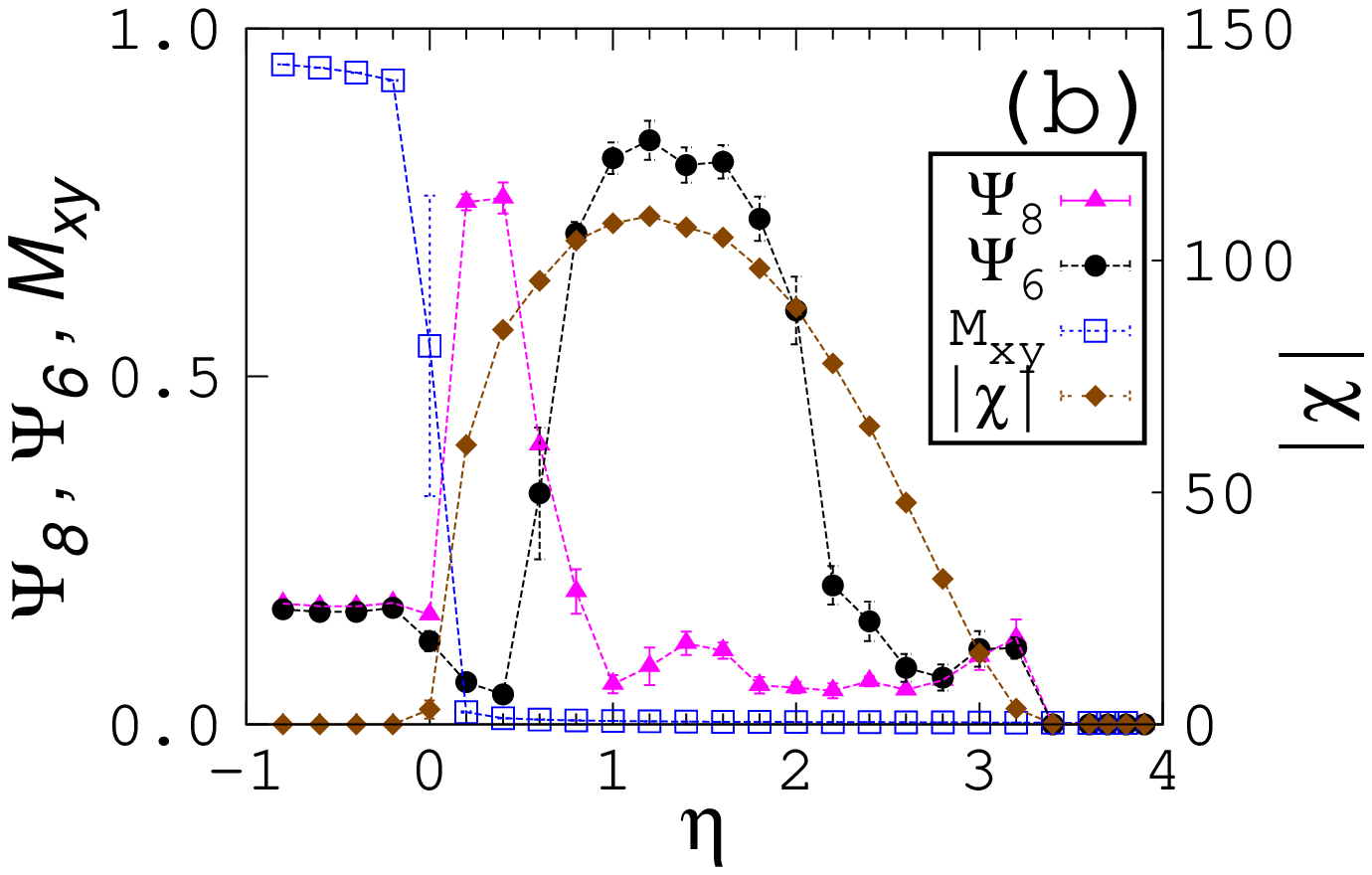} 
\end{center}
\caption{$\eta$ dependences of the order parameters at (a) $H=0$ and (b) $H=1$ with 
the weak dipolar interaction $D'=0.3$ at $T=0.1$ for $L=84$.}
\label{O_dp03}
\end{figure}

\begin{figure}
\begin{center}
\includegraphics[width = 7.25cm]{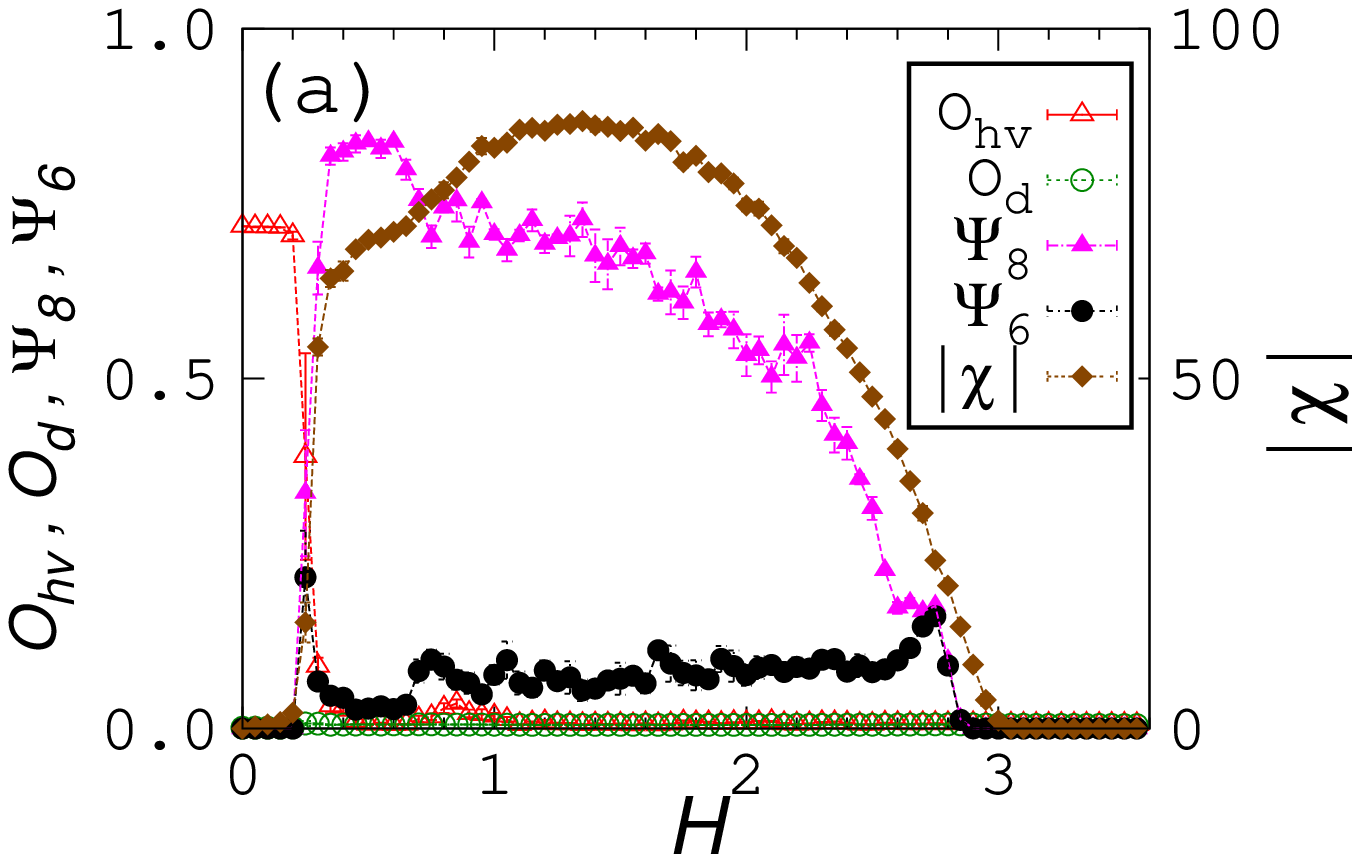} 
\includegraphics[width = 7.25cm]{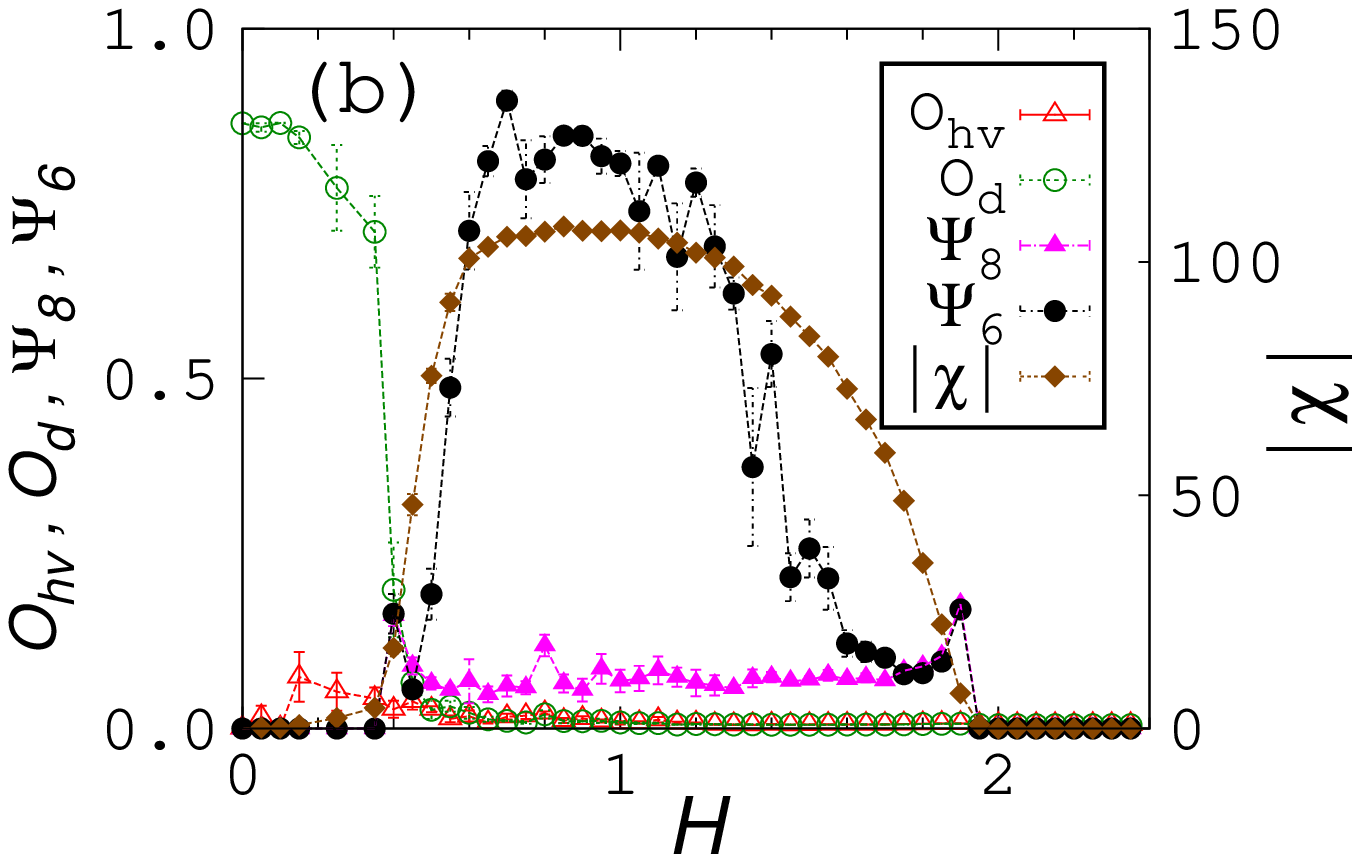} 
\end{center}
\caption{$H$ dependences of the order parameters at (a) $\eta=0.4$ and (b)  $\eta=1.5$ with the weak dipolar interaction $D'=0.3$ at $T=0.1$ for $L=84$. }
\label{O_dp03_Hdep}
\end{figure}

\subsection{The system with the weak dipolar interaction}
\label{weak-dipole}

We study the model with the weak dipolar interaction. 
The $\eta$ dependences of the order parameters for $D'=0.3$ at $H=0$ and $H=1$ are displayed in Figs.~\ref{O_dp03} (a) and \ref{O_dp03} (b), respectively (see also Fig.~\ref{PD} (b)). 
The $H$ dependences of the order parameters for $D'=0.3$ at $\eta=0.4$ and $\eta=1.5$ are shown in Figs.~\ref{O_dp03_Hdep} (a) and \ref{O_dp03_Hdep} (b), respectively.

With the dipolar interaction, the region of the canted ferromagnetic phase expands to the positive $\eta$ direction, and the regions of the stripe and skyrmion-lattice phases are shifted to this direction. 
Furthermore, the dipolar interaction expands the region of the stripe and skyrmion-lattice phases in both $\eta$ and $H$ directions. 
Namely, the dipolar interaction stabilizes the ordered states.

It should be noted that unlike the case without the dipolar interaction, the diagonal square skyrmion-lattice phase does not appear, and instead the upright square skyrmion-lattice phase appears between the canted ferromagnetic and triangular skyrmion-lattice phases, and any skyrmion phases do not appear at zero and low fields. 

With increasing $\eta$ at low fields, the canted ferromagnetic phase changes to the vertical helical phase, to the diagonal helical phase, and to the vertical helical phase again. Namely, the direction of the stripe changes twice. We analyze the cause of this change of the stripe direction in Sec.~\ref{helical}. 
These findings lead to four other types of triple points II-V at $(\eta,H) \simeq (0.3,0.2)$, $(\eta,H) \simeq (0.7,0.38)$, $(\eta,H) \simeq (1.1,0.42)$, $(\eta,H) \simeq (2.9,0.35)$, respectively, which are marked with open squares in Fig.~\ref{PD} (b). Triple point II is that of the canted ferromagnetic, upright square skyrmion-lattice, and vertical helical phases, triple point III is that of upright square skyrmion-lattice, triangular skyrmion-lattice, and vertical helical phases, and triple points IV and V are those of triangular skyrmion-lattice, vertical helical, and diagonal helical phases.

At large $\eta$, the spins are Ising-like and the anisotropy term is more important than the DM term (vector product of spins approaches zero). 
The vertical helical phase shows an Ising-like stripe pattern and is stabilized again for $\eta \gtrsim 3.0$. 
Vertical stripe patterns are a characteristic of the dipolar Ising model~\cite{Booth,MacIsaac-Ising,Toloza,Rastelli06,Cannas,Pighin-Ising,Rastelli,Vindigni,Rizzi,Fonseca,Ruger,Horowitz,Bab,Leib,Komatsu1}. 

\subsection{The system with the strong dipolar interaction}
\label{strong-dipole}
\begin{figure}
\begin{center}
\includegraphics[width = 6.5cm]{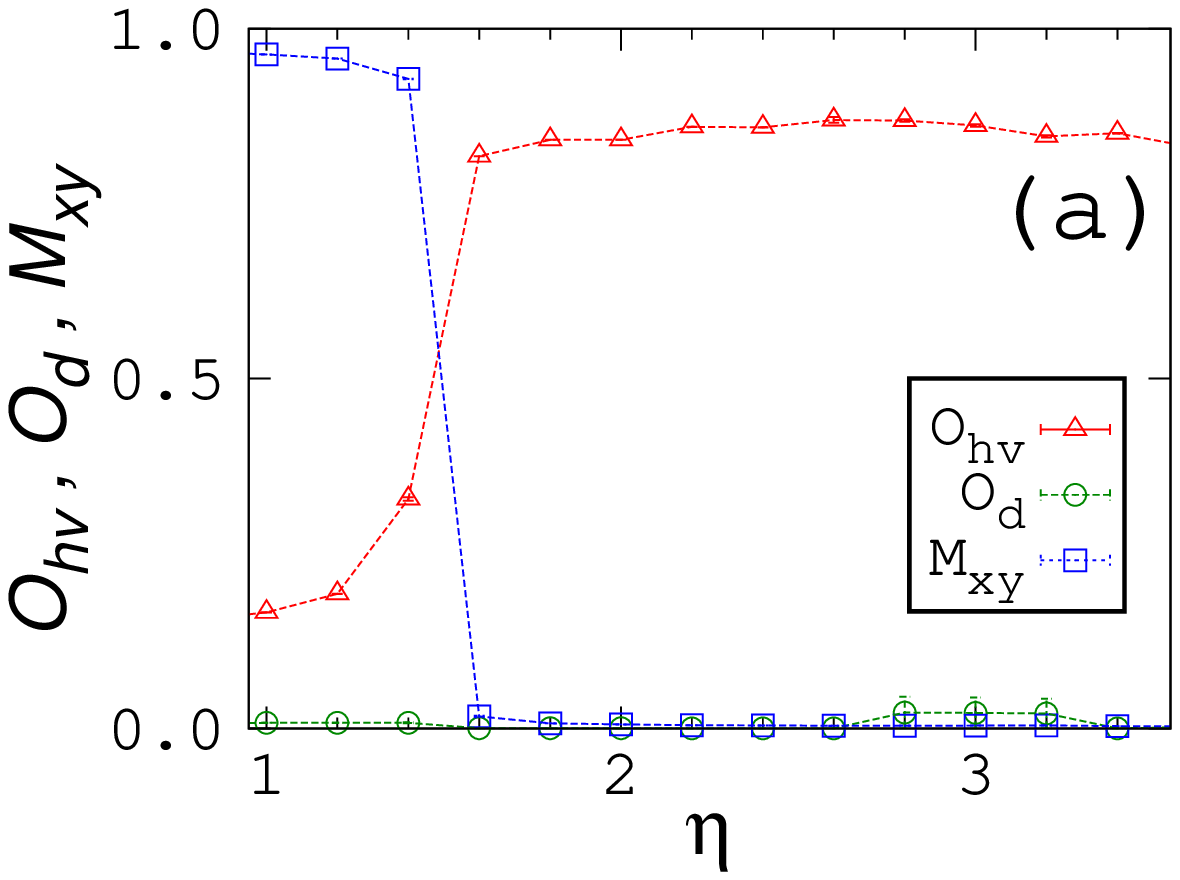} 
\hspace*{0.6cm}
\includegraphics[width = 7.25cm]{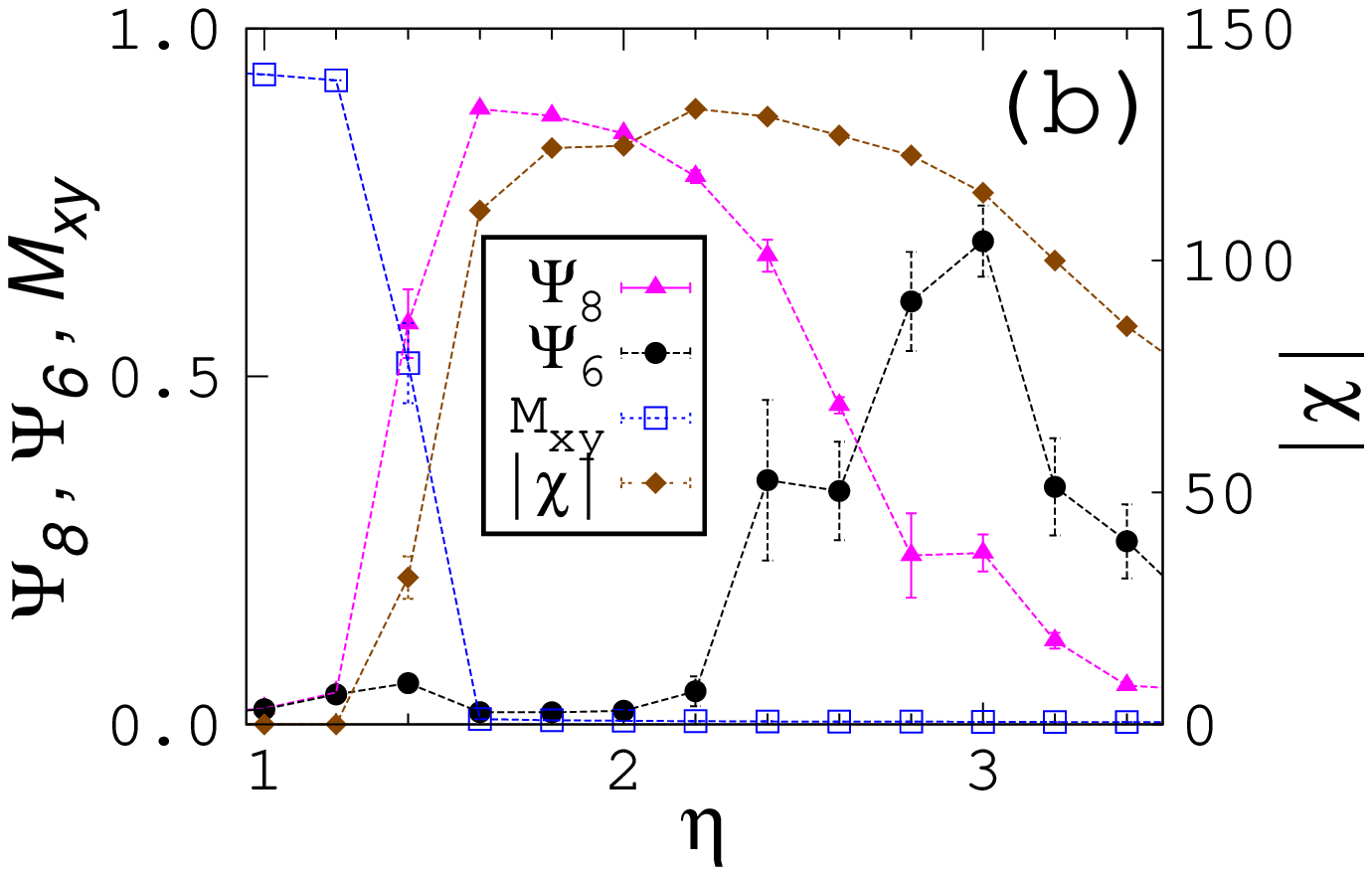} 
\end{center}
\caption{$\eta$ dependences of the order parameters at (a) $H=0$ and (b) $H=1.5$ with the strong dipolar interaction $D'=0.6$ at $T=0.1$ for $L=84$.}
\label{O_dp06}
\end{figure}

\begin{figure}
\begin{center}
\includegraphics[width = 7.25cm]{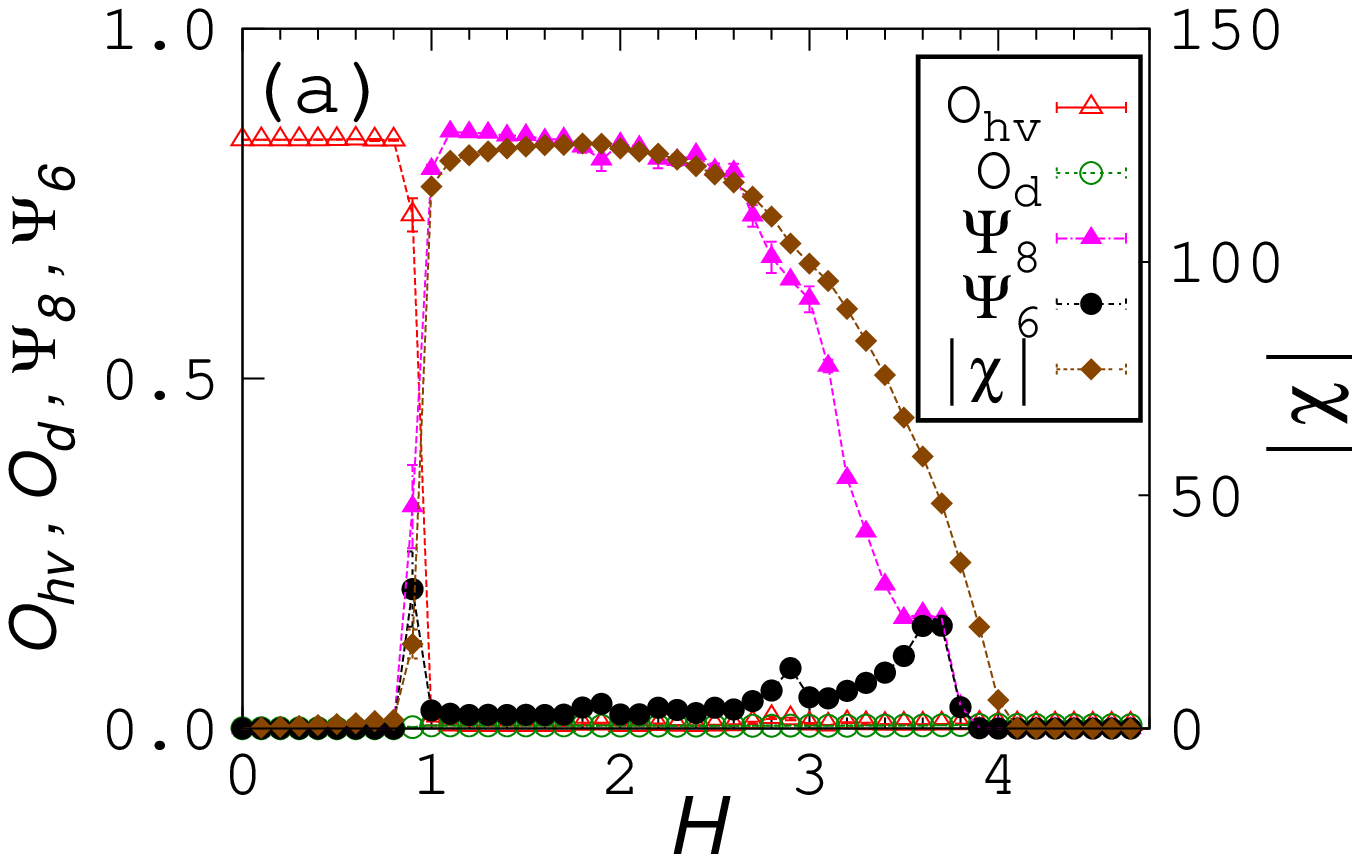} 
\includegraphics[width = 7.25cm]{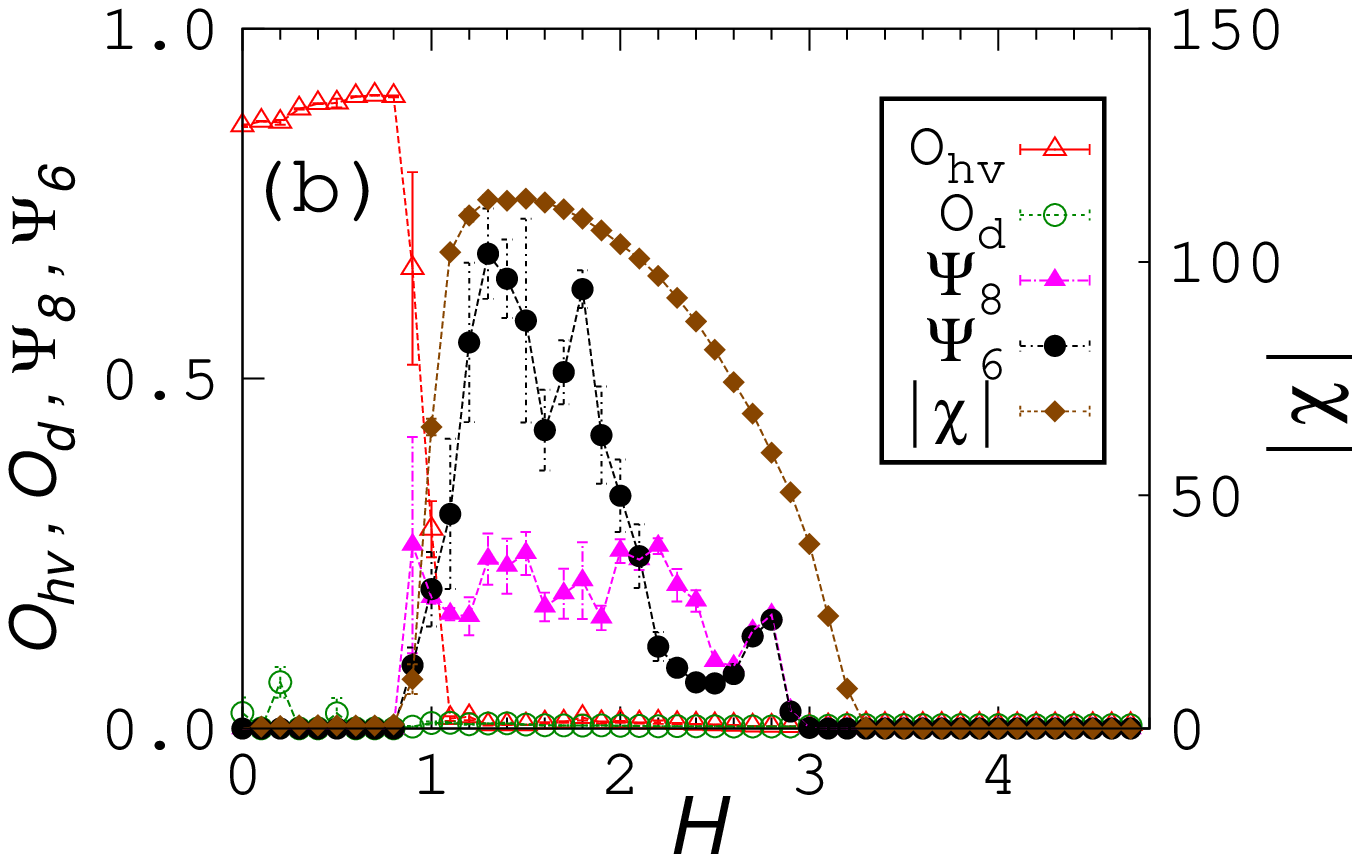} 
\end{center}
\caption{$H$ dependences of the order parameters at (a) $\eta=2$ and (b)  $\eta=3$ with the strong dipolar interaction $D'=0.6$ at $T=0.1$ for $L=84$.}
\label{O_dp06_Hdep}
\end{figure}

We investigate the effect of the strong dipolar interaction.  
The $\eta$ dependences of the order parameters for $D'=0.6$ at $H=0$ and $H=1.5$ are displayed in Figs.~\ref{O_dp06} (a) and \ref{O_dp06} (b), respectively (see also Fig.~\ref{PD} (c)). The $H$ dependences of the order parameters for $D'=0.6$ at $\eta=2$ and $\eta=3$ are given in Figs.~\ref{O_dp06_Hdep} (a) and \ref{O_dp06_Hdep} (b), respectively.

Due to the strong dipolar interaction, the canted ferromagnetic phase further expands to the positive $\eta$ direction. The upright square skyrmion and triangular skyrmion-lattice phases appear in the same manner as $D'=0.3$, but their regions expand to the positive $\eta$ and $H$ directions. 

We find that unlike $D'=0$ and $D'=0.3$, only a vertical stripe appears in the helical phase, which is stabilized in a wider region in the $H-\eta$ space.  
In this case, there exist two triple points VI and VII at $(\eta,H) \simeq (1.45, 0.74)$ and $(\eta,H) \simeq (2.7, 1.02)$, respectively, which are marked with open squares in Fig.~\ref{PD} (c). These are the same types as points II and III, respectively.  

We also find that the peak value of $|\chi|$ becomes larger for larger $D'$. 
Since $|\chi|$ is proportional to the number of the skyrmions, 
the peak value of the density of skyrmions increases for stronger dipolar interaction.  
Indeed we observe that the size of skyrmions is smaller and the density is higher for larger $D'$ in the snapshots of the skyrmion-lattice phases in Fig.~\ref{snap1} and Figs.~\ref{snap2} (a), \ref{snap2} (b), and \ref{snap2} (c). 
We also observe that the stripe width (pitch length) in the helical ordered phase reduces for larger $D'$, which is similar to the shortening of the stripe width in the stripe-ordered phase in the 2D Ising dipolar model~\cite{Pighin}. 

%

\subsection{The ground state energies in the helical phases}
\label{helical}

For $D'=0.3$, we observed that the stripe direction in the helical phases depends on $\eta$. 
Here we clarify the origin of this behavior. 
We evaluate the ground state energies of the two helical phases 
at $H=0$ using variational functions for spins. 

The variational function is defined as 
\begin{equation}
\vector{S} = \frac{ \vector{S} _0 }{ | \vector{S} _0 | }. 
\label{S_test}
\end{equation}
In the vertical helical phase, $\vector{S} _0$ is expressed as    
\begin{eqnarray}
S_{0, (i_x , i_y)} ^x & = & 0, \\
S_{0, (i_x , i_y)} ^y & = & -\sin \left\{ k(i_x + \phi_0 )  \right\}, \\
{\rm and \;\;\;} S_{0, (i_x , i_y)} ^z & = & \alpha \cos \left\{ k(i_x + \phi_0 )  \right\} ,
\label{S0_str}
\end{eqnarray}
and in the diagonal helical phase, $\vector{S} _0$ is described by 
\begin{eqnarray}
S_{0, (i_x , i_y)} ^x & = & \sin \left\{ k(i_x + i_y + \phi_0 )  \right\}, \\
S_{0, (i_x , i_y)} ^y & = & -\sin \left\{ k(i_x + i_y + \phi_0 )  \right\}, \\
{\rm and \;\;\;}  S_{0, (i_x , i_y)} ^z & = & \alpha \cos \left\{ k(i_x + i_y + \phi_0 )  \right\} ,
\label{S0_diagonal}
\end{eqnarray}
where $k$, $\phi_0$, and $\alpha$ are the variational parameters. Variational functions based on the sinusoidal functions have been used in order to evaluate the energy in the helical phases in isotropic systems~\cite{Kwon}. 
In the present study, we take the Ising anisotropy $\eta$ into account in the model, and to include this effect, the parameter $\alpha$ is introduced. 

\begin{figure}
\begin{center}
\includegraphics[width = 6.5cm]{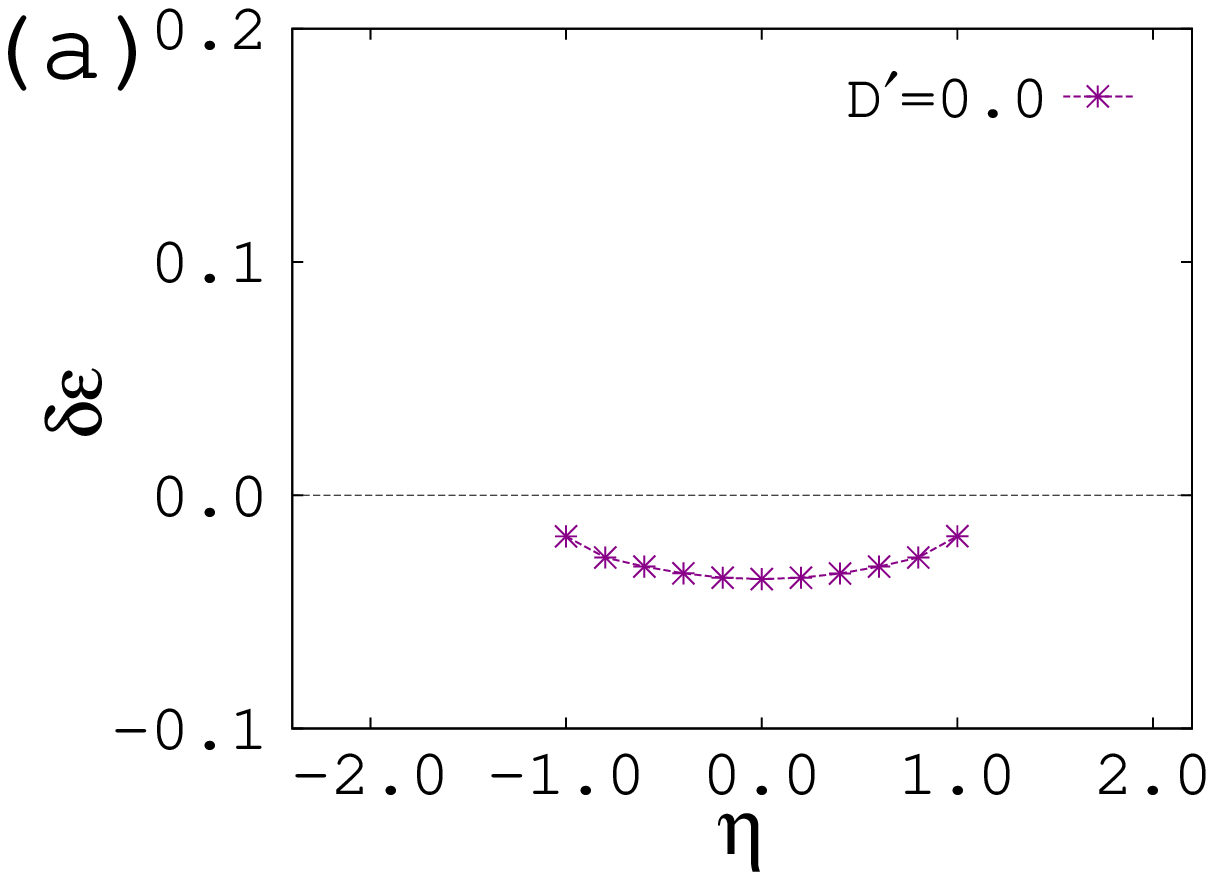} 
\includegraphics[width = 6.5cm]{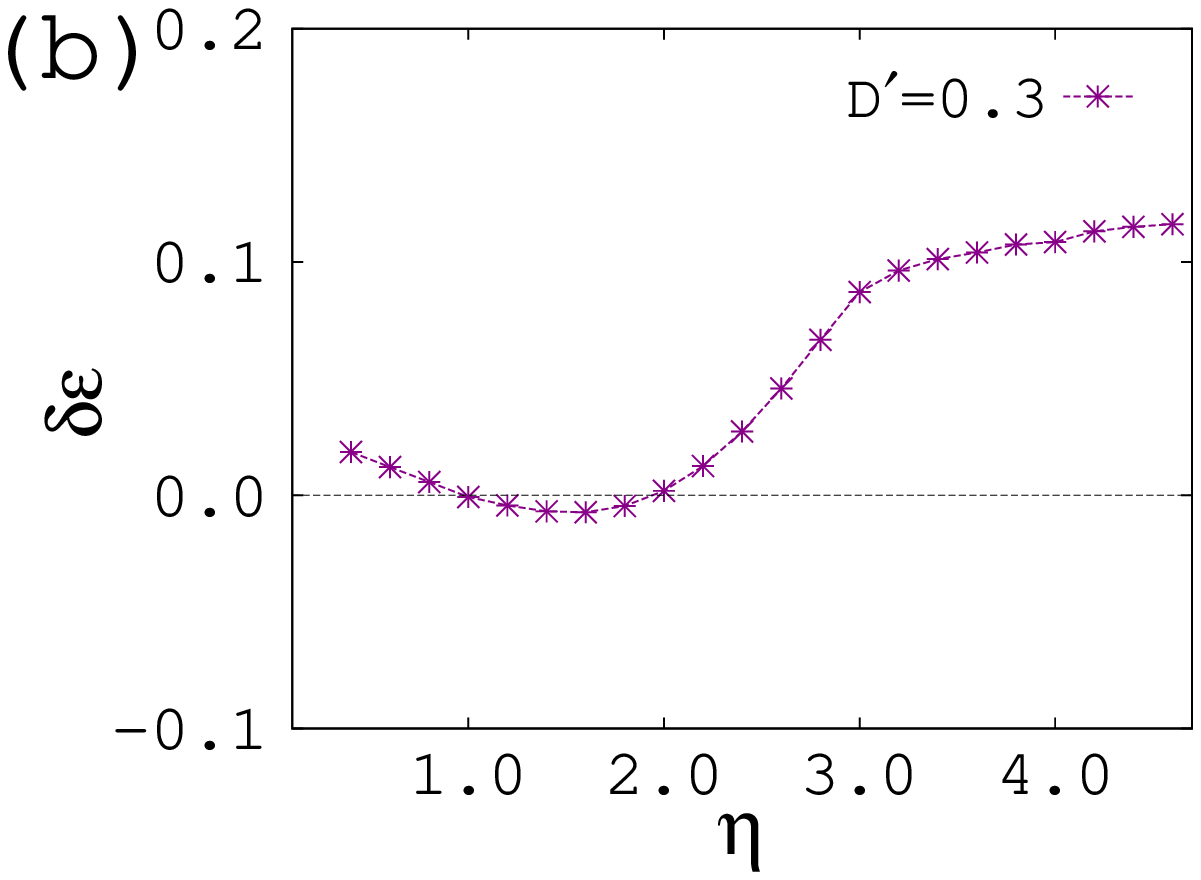} 
\includegraphics[width = 6.5cm]{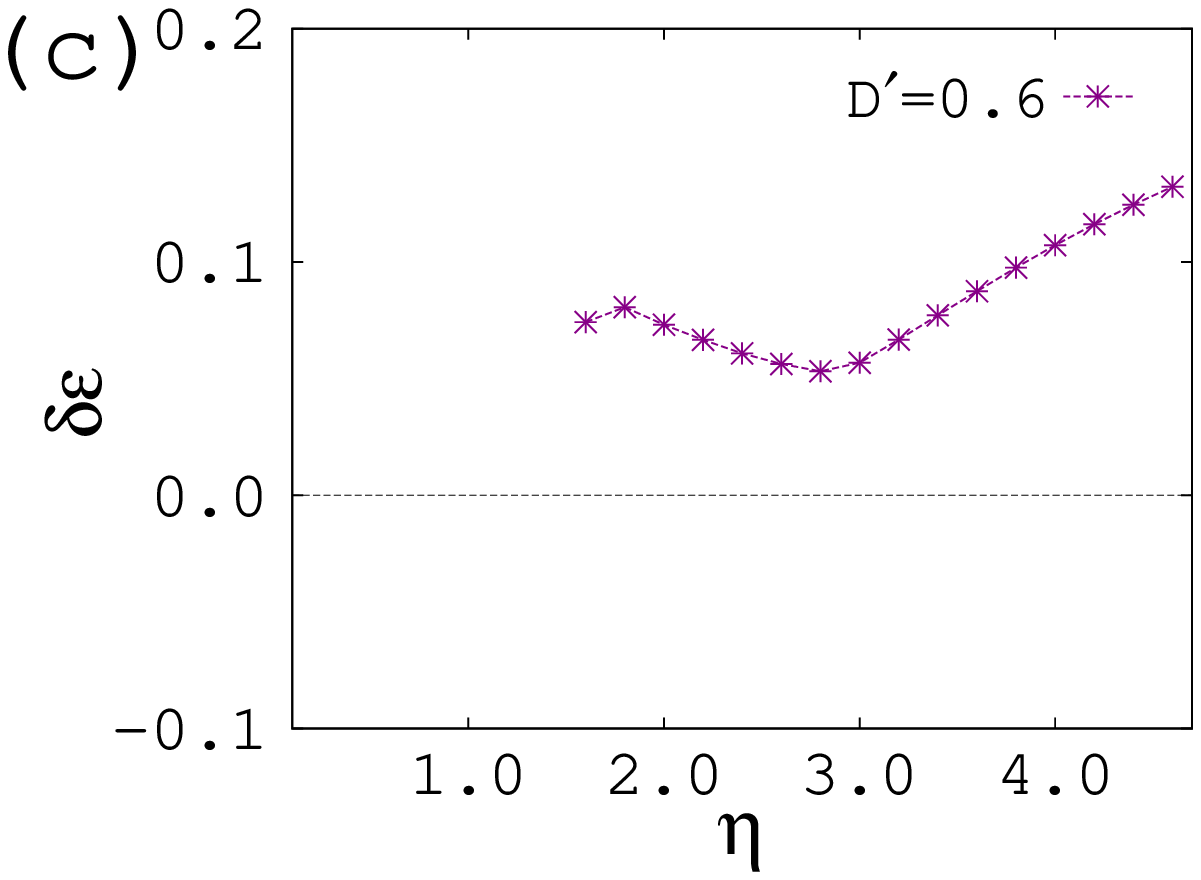} 
\end{center}
\caption{ $\eta$ dependences of $\delta \epsilon$ (Eq. (\ref{delta_e})) for (a) $D'=0$, (b) $D'=0.3$, and (c) $D'=0.6$. The vertical and diagonal helical phases are stable for $\delta \epsilon > 0$ and $\delta \epsilon < 0$, respectively.}
\label{dE_eval}
\end{figure}
We calculate the ground state energies of the two helical phases $E_{\mathrm{diagonal}}$ and $E_{\mathrm{vertical}}$, and plot their difference per spin, 
\begin{equation}
\delta \epsilon \equiv \frac{1}{N} (E_{\mathrm{diagonal}} - E_{\mathrm{vertical}} )
\label{delta_e}
\end{equation}
 as a function of $\eta$ in Fig.~\ref{dE_eval}. We exclude the points at which the energy of the phase with uniform in-plane magnetization $E_{\rm uniform}$ satisfies $E_{\rm uniform} < E_{\rm diagonal}$ and $E_{\rm uniform} < E_{\rm vertical}$. 
For the calculation of $E_{\mathrm{uniform} }$,  we use the condition that all the spins have the same direction.

For $D'= 0.0$, the diagonal helical phase is stable ($\delta \epsilon < 0$) (Fig.~\ref{dE_eval} (a)), while for $D'= 0.6$ the vertical helical phase is stable ($\delta \epsilon > 0$) (Fig.~\ref{dE_eval} (c)). 
For $D' = 0.3$, the diagonal helical phase is stable for $1.0 \leq \eta \leq 2.0$, and the vertical one is stable for $\eta \leq 1.0$ or $2.0 \leq \eta$ (Fig.~\ref{dE_eval} (b)). 
Namely, the stability of the helical phase changes between the diagonal and vertical structures depending on $\eta$, which explains qualitatively the observation in Sec.~\ref{weak-dipole}, i.e., with increasing $\eta$, the vertical helical phase changes to the diagonal helical phase and to the vertical helical phase again. 

For the skyrmion-lattice phases, we find that more variational parameters are necessary and it is difficult to analyze the stability by this approach.

\section{Summary \label{summary} }

We investigated the two-dimensional classical Heisenberg model with the ferromagnetic exchange ($J$), Dzyaloshinskii-Moriya ($D$) and dipolar ($D'$) interactions, and the Ising anisotropy ($\eta$) with $J=D=1$. 
We estimated the order parameters for square and triangular skyrmion-lattice phases, diagonal and vertical helical phases, and in-plane magnetization 
with the stochastic cut-off $O(N)$ Monte Carlo method. We presented the field ($H$) vs.\ $\eta$ phase diagrams at a low temperature $T=0.1$ for three typical values of the dipolar interaction, $D' = 0.0$, $0.3$, and $0.6$ in Fig.~\ref{PD}. 

When there is no dipolar interaction ($D'=0$), for small $D$ compared to $J$, any skyrmion-lattice phases do not exist at zero and low fields, and the triangular skyrmion-lattice phase appears at high fields. 
However, for $D$ close to $J$ ($D = J =1$ in the present study), the diagonal square skyrmion-lattice phase exists at zero and low fields between the canted ferromagnetic phase at smaller $\eta$ and the diagonal helical phases at larger $\eta$. At high fields the triangular skyrmion-lattice phase appears at larger $\eta$ than the diagonal square skyrmion-lattice phase. There exists a triple point at which the diagonal square skyrmion-lattice, triangular skyrmion-lattice, and diagonal helical phases coexist. 

The dipolar interaction leads to other types of skyrmion lattice and helical phases, 
which yield four other types of triple points. 
The effect of the dipolar interaction shifts the phase boundaries to the positive $\eta$ and $H$ directions and stabilizes the ordered phases, whose regions in the $H-\eta$ phase diagram are expanded. 
The dipolar interaction also increases the skyrmion density and reduces the skyrmion size in the skyrmion-lattice phase and the pitch length (stripe width) in the helical phase.

In both cases of the weak ($D'=0.3$) and strong ($D'=0.6$) dipolar interactions, any skyrmion-lattice phases do not exist at zero and low fields, which is different from the case without the dipolar interaction ($D'=0$). 
For the weak dipolar interaction ($D'=0.3$) at high fields, instead of the diagonal square skyrmion-lattice phase, the upright square skyrmion-lattice phase appears between the canted ferromagnetic phase at smaller $\eta$ and the triangular skyrmion-lattice phase at larger $\eta$. 
With increasing $\eta$ at zero and low fields, the canted ferromagnetic phase changes to the vertical helical phase, to the diagonal helical phase, and to the vertical helical phase again. This reentrant transition is qualitatively explained by using the variational functions for spins. For the strong dipolar interaction ($D'=0.6$), at zero and low fields, only the vertical helical structure exists at larger $\eta$ than the canted ferromagnetic phase. 

In the present paper, we focused on the model for mterials with non-centrosymmetric lattice structures, in which the DM interaction plays an important role. 
Very recently, a square skyrmion-lattice phase was discovered in GdRu$_2$Si$_2$~\cite{Khanh}, which has a centrosymmetric structure. There the DM interaction is not essential, and four-spin interactions are considered to be important. 
The present paper shows a possible scenario of square skyrmion-lattice phases driven by the DM interaction with or without the dipolar interaction.

\begin{acknowledgments}
The present study was supported by Grants-in-Aid for Scientific Research C (No. 18K03444 and No. 20K03809) from the Ministry of Education, Culture, Sports, Science and Technology (MEXT) of Japan, and the Elements Strategy Initiative Center for Magnetic Materials (ESICMM) (Grant No. 12016013) funded by MEXT. 
The calculations were partially performed using the supercomputer at the Supercomputer Center of the Institute for Solid State Physics, the University of Tokyo, and the Numerical Materials Simulator at the National Institute for Materials Science. 
\end{acknowledgments}


\begin{thebibliography}{99}


\bibitem{Heinrich}
B. Heinrich, J.A.C. Bland (Eds.), Ultrathin Magnetic Structures, Springer-Verlag, Berlin, 2004.

\bibitem{Bader}
S. D. Bader, Rev. Mod. Phys {\bf 78}, 1, (2006).

\bibitem{Pappas}
D.~P.~Pappas, K.-P.~K\"amper, and H.~Hopster, Phys.\ Rev.\ Lett.\ {\bf 64}, 3179 (1990).
\bibitem{Allenspach}
R.~Allenspach and A.~Bischof, Phys.\ Rev.\ Lett.\ {\bf 69}, 3385 (1992). 

\bibitem{Ramchal}
R.~Ramchal, A.~K.~Schmid, M.~Farle, and H.~Poppa, Phys.\ Rev.\ B {\bf 69}, 214401 (2004).

\bibitem{Won}
C. Won, Y. Z. Wu, J. Choi, W. Kim, A. Scholl, A. Doran, T. Owens, J. Wu, X. F. Jin, H. W. Zhao, and Z. Q. Qiu, Phys.\ Rev.\ B {\bf 71}, 224429 (2005). 

\bibitem{Qiu}
Z.~Q.~Qiu, J.~Pearson, and S.~D.~Bader, Phys.\ Rev.\ Lett.\ {\bf 70}, 1006 (1993).

\bibitem{Booth} I.~Booth, A.~B.~MacIsaac, J.~P.~Whitehead, and K.~De'Bell, 
Phys.\ Rev.\ Lett.\ \textbf{75}, 950 (1995).

\bibitem{MacIsaac-Ising}  A.~B.~MacIsaac, J.~P.~Whitehead, M.~C.~Robinson, 
and K.~De'Bell, Phys.\ Rev.\ B {\textbf 51}, 16033 (1995).

\bibitem{Toloza} 
J.~H.~Toloza, F.~A.~Tamarit, and S.~A.~Cannas, Phys.\ Rev.\ B \textbf{58}, R8885 (1998).

\bibitem{Rastelli06} E.~Rastelli, S.~Regina, and A.~Tassi, Phys.\ Rev.\ B \textbf{73}, 144418 (2006).

\bibitem{Cannas} S.~A.~Cannas, M.~F.~Michelon, D.~A.~Stariolo, and F.~A.~Tamarit, 
Phys.\ Rev.\ B \textbf{73}, 184425 (2006).

\bibitem{Pighin-Ising} S.~A.~Pighin and S.~A.~Cannas Phys.\ Rev.\ B \textbf{75}, 224433 (2007). 

\bibitem{Rastelli} E.~Rastelli, S.~Regina, and A.~Tassi, Phys.\ Rev.\ B \textbf{76}, 054438 (2007).

\bibitem{Vindigni} A.~Vindigni, N.~Saratz, O.~Portmann, D.~Pescia, and P.~Politi, 
Phys.\ Rev.\ B \textbf{77}, 092414 (2008).

\bibitem{Rizzi} L.~G.~Rizzi and N.~A.~Alves, Physica B \textbf{405}, 1571 (2010).

\bibitem{Fonseca} J.~S.~M.~Fonseca, L.~G.~Rizzi, and N.~A.~Alves, 
Phys.\ Rev.\ E \textbf{86}, 011103 (2012).

\bibitem{Ruger} R.~R\"{u}ger and R.~Valent\'{i}, Phys.\ Rev.\ B \textbf{86}, 024431 (2012). 

\bibitem{Horowitz} C.~M.~Horowitz, M.~A.~Bab, M.~Mazzini, M.~L.~Rubio Puzzo, 
and G.~P.~Saracco, Phys.\ Rev.\ E \textbf{92}, 042127 (2015).

\bibitem{Bab} M.~A.~Bab, C.~M.~Horowitz, M.~L.~Rubio Puzzo, and G.~P.~Saracco, 
Phys.\ Rev.\ E \textbf{94}, 042104 (2016).


\bibitem{Leib}
The ground state of the 2D dipolar Ising ferromagnet for large $J$ was proved to be a periodic stripe state in the literature: A. Giuliani, J. L. Lebowitz, and E. H. Lieb, Phys.\ Rev.\ B {\bf 74}, 064420 (2006). 

\bibitem{Komatsu1}
H.~Komatsu, Y.~Nonomura, and M.~Nishino, Phys.\ Rev.\ E \textbf{98}, 062126 (2018). 

\bibitem{Pescia}
D.~Pescia and V.~L.~Pokrovsky, Phys.\ Rev.\ Lett.\ {\bf 65}, 2599 (1990).

\bibitem{Moschel}
A.~Moschel and K.~D.~Usadel, J.\ Magn.\ Magn.\ Mater.\ {\bf 140-144}, 649 (1995).

\bibitem{Hucht}
A.~Hucht, A.~Moschel, K.~D.~Usadel, J.\ Magn.\ Magn.\ Mater.\ {\bf 148}, 32 (1995).

\bibitem{MacIsaac1}
A.~B.~MacIsaac, J.~P.~Whitehead, K.~De'Bell, and P.~H.~Poole, Phys.\ Rev.\ Lett.\ {\bf 77}, 739 (1996).

\bibitem{MacIsaac2}
A. B. MacIsaac, K. De'Bell, and J. P. Whitehead, Phys.\ Rev.\ Lett.\ {\bf 80}, 616 (1998). 

\bibitem{Bell}
K. De'Bell, A. B. MacIsaac, and  J. P. Whitehead,  
Rev. Mod. Phys. {\bf 72}, 225 (2000). 

\bibitem{Santamaria}
C. Santamaria and H.T. Diep, J. Magn. Magn. Mater. {\bf 212}, 23 (2000).

\bibitem{Rapini} 
M. Rapini, R. A. Dias, and B. V. Costa, Phys. Rev. B {\bf 75}, 014425 (2007).

\bibitem{Whitehead}
J. P. Whitehead, A. B. MacIsaac, and K. De'Bell, Phys. Rev. B {\bf 77}, 174415 (2008).

\bibitem{Carubelli} M.~Carubelli, O.~V.~Billoni, S.~A.~Pighin, S.~A.~Cannas, D.~A.~Stariolo, and F.~A.~Tamarit, 
Phys.\ Rev.\ B \textbf{77}, 134417 (2008).

\bibitem{Mol}
L. A. S. M\'ol and B. V. Costa, Phys. Rev. B {\bf 79}, 054404 (2009).

\bibitem{Pighin2}
S. A. Pighin, O. V. Billoni, D. A. Stariolo, and S. A. Cannas,
J. Magn. Magn. Mater. {\bf 322}, 3889 (2010).

\bibitem{Pighin}
S. A. Pighin, O. V. Billoni, and S. A. Cannas, Phys.\ Rev.\ B {\bf 86}, 051119 (2012).

\bibitem{Mol2}
L. A. S. M\'ol and B. V. Costa, J. Magn. Magn. Mater. {\bf 353}, 11 (2014).

\bibitem{Komatsu2}
H.~Komatsu, Y~Nonomura, and M.~Nishino, Phys.\ Rev.\ B \textbf{100}, 094407 (2019).



\bibitem{Muhlbauer}
 S. M\"uhlbauer, B. Binz, F. Jonietz, C. Pfleiderer, A. Rosch, A. Neubauer, R. Georgii, and P. B\"oni, Science {\bf 323} 915 (2009). 

\bibitem{Yu}
X. Z. Yu, Y. Onose, N. Kanazawa, J. H. Park, J. H. Han, Y. Matsui, N. Nagaosa, and Y. Tokura, Nature {\bf 465}, 901 (2010).

\bibitem{Jonietz11} F.~Jonietz, S.~M\"{u}hlbauer, C.~Pfleiderer, A.~Neubauer, W.~M\"{u}nzer, A.~Bauer, T.~Adams, R.~Georgii, P.~B\"{o}ni, R.~A.~Duine, K.~Everschor, M.~Garst, and A. Rosch, Science \textbf{330}, 1648 (2010).

\bibitem{Yu11} X.~Z.~Yu, N.~Kanazawa, Y.~Onose, K.~Kimoto, W.~Z.~Zhang, S.~Ishiwata, Y.~Matsui, and Y.~Tokura, Nat.\ Mater.\ \textbf{10}, 106 (2011).

\bibitem{Schulz12} T.~Schulz, R.~Ritz, A.~Bauer, M.~Halder, M.~Wagner, C.~Franz, C.~Pfleiderer, K.~Everschor, M.~Garst, and A.~Rosch, Nat.\ Phys.\ \textbf{8} 301 (2012).



\bibitem{Sampaio13} J.~Sampaio, V.~Cros, S.~Rohart, A.~Thiaville, and A.~Fert, Nat.\ Nanotechnol.\ \textbf{8}, 839 (2013).


\bibitem{Jiang15} W.~Jiang, P.~Upadhyaya, W.~Zhang, G.~Yu, M.~B.~Jungfleisch, F.~Y.~Fradin, J.~E.~Pearson, Y.~Tserkovnyak, K.~L.~Wang, O.~Heinonen, S.~G.~E.~te~Velthuis, and A.~Hoffmann, Science \textbf{349}, 283 (2015).

\bibitem{Boulle16} O.~Boulle, J.~Vogel, H.~Yang, S.~Pizzini, D.~de~Souza.~Chaves, A.~Locatelli, T.~O.~Mente\c{s}, A.~Sala, L.~D.~Buda-Prejbeanu, O.~Klein, M.~Belmeguenai, Y.~Roussign\'{e}, A.~Stashkevich, S.~M.~Ch\'{e}rif, L.~Aballe, M.~Foerster, M.~Chshiev, S.~Auffret, I.~M.~Miron, and G.~Gaudin, Nat.\ Nanotechnol.\ \textbf{11}, 449 (2016).



\bibitem{Yi} S.~D.~Yi, S.~Onoda, N.~Nagaosa, and J.~H.~Han,
Phys.\ Rev.\ B \textbf{80}, 054416 (2009).

\bibitem{Kwon} H.~Y.~Kwon, K.~M.~Bu , Y.~Z.~Wu, and C.~Won,
J.\ Mag.\ Mag.\ Mat.\ \textbf{324} 2171 (2012). 


\bibitem{Banerjee} S. Banerjee, J. Rowland, O. Erten, and M. Randeria, Phys.\ Rev.\ X \textbf{4}, 031045 (2014).

\bibitem{Lin} S.~Z.~Lin, A.~Saxena, and C.~D.~Batista, Phys.\ Rev.\ B \textbf{91}, 224407 (2015).

\bibitem{Rowland}
J. Rowland, S. Banerjee, and M. Randeria,  Phys.\ Rev.\ B  {\bf 93}, 020404(R) (2016). 

\bibitem{Nishikawa}
Y. Nishikawa, K. Hukushima, and W. Krauth,  Phys.\ Rev.\ B \textbf{99}, 064435 (2019).

\bibitem{Bernard} A.~Bernard-Mantel, C.~B.~Muratov, and T.~M.~Simon,
Phys.\ Rev.\ B \textbf{101}, 045416 (2020).



\bibitem{Sasaki} M.~Sasaki and F.~Matsubara, J.\ Phys.\ Soc.\ Jpn.\ \textbf{77}, 024004 (2008). 


\bibitem{Shewchuk}
J. R. Shewchuk, Computational Geometry {\bf 22}, 21 (2002).

\bibitem{Lenov}
See A. O. Leonov, T. L. Monchesky, N. Romming, A. Kubetzka, A. N. Bogdanov and R. Wiesendanger, New J. Phys. {\bf 18}, 065003 (2016) for a detailed discussion about the boundary between skyrmion-lattice and polarized ferromagnetic phases. 

\bibitem{Kashuba}
A. B. Kashuba and V. L. Pokrovsky, Phys.\ Rev.\ B {\bf 48}, 10335 (1993). 



\bibitem{Khanh}
N. D. Khanh et al., Nat. Nanotechnol. {\bf 15}, 444 (2020). 


\end{thebibliography}
\end{document}